# Beyond Olfaction: New Insights into Human Odorant Binding Proteins

**Mifen Chen[a], Soufyan Lakbir[a][b], Mihyeon Jeon[a], Vojta Mazur[a], Sanne Abeln[b], and Halima Mouhib[a]**

[a] Department of Computer Science, VU Bioinformatics Group, Vrije Universiteit Amsterdam, De Boelelaan 1105, 1081 HV Amsterdam, The Netherlands.
[b] Department of Computer Science, AI Technology for Life, Universiteit Utrecht, Heidelberglaan 8, 3584 CS Utrecht, The Netherlands

## Abstract

Until today, the exact function of mammalian odorant binding proteins (OBPs) remains a topic of debate. Although their main established function lacks direct evidence in human olfaction, OBPs are traditionally believed to act as odorant transporters in the olfactory sense, which led to the exploration of OBPs as biomimetic sensor units in artificial noses. Now, available RNA-seq and proteomics data identified the expression of human OBPs (hOBP2A and hOBP2B) in both, male and female reproductive tissues. This observation prompted the conjecture that OBPs may possess functions that go beyond the olfactory sense, potentially as hormone transporters. Such a function could further link them to the tumorigenesis and cancer progression of hormone dependent cancer types including ovarian, breast, prostate and uterine cancer. In this structured review, we use available data to explore the effects of genetic alterations such as somatic copy number aberrations and single nucleotide variants on OBP function and their corresponding gene expression profiles. Our computational analyses suggest that somatic copy number aberrations in OBPs are associated with large changes in gene expression in reproductive cancers while point mutations have little to no effect. Additionally, the structural characteristics of OBPs, together with other lipocalin family members, allow us to explore putative functions within the context of cancer biology. Our overview consolidates current knowledge on putative human OBP functions, their expression patterns, and structural features. Finally, it provides an overview on applications, highlighting emerging hypotheses and future research directions within olfactory and non-olfactory roles.

## Introduction

Since their first discovery in the olfactory mucus of cows (Bignetti et al., 1985, Pelosi et al., 1982), mammalian odorant binding proteins (OBPs) are believed to play a role in the initial stages of olfactory perception (Lacazette, 2000, Lazar et al., 2002; Grolli et al., 2006). Due to their occurrence in the nose, the subsequent studies mainly focussed on understanding their role in the olfactory sense (Melis et al., 2021), where they are now recognized as players in peri-receptor events (Heydel et al., 2013). Hereby, OBPs are thought to bind and transport hydrophobic odorant molecules by lowering the energy barrier through the aqueous environment to olfactory receptors (Paesani et al., 2025). They possess a strong hydrophobic calyx that can facilitate the binding of hydrophobic molecules like odorants. The typical beta-barrel structure and the attached C-terminal alpha-helix exhibit a conserved structural feature across all mammalian OBPs, which suggests similar functions throughout different species (Pelosi and Knoll, 2022). While mammals possess hundreds of different olfactory receptors (ORs) which are able to distinguish between thousands of volatile odorant molecules, only a small number of OBPs, usually 2-3, were identified in the mammalian nasal mucus. Even with the exception of porcupines with eight OBPs (Pelosi and Knoll, 2022, Ganni et al., 1997, Dal Monte et al., 1991), the number is significantly smaller than the ORs and suggests that OBPs are mostly non-specific binders that can transport a wide range of structurally different odorants and volatiles (Pelosi and Knoll, 2022). This complies with a putative function as transporter proteins, although it has been shown that their specificity may be enhanced through selected mutagenesis experiments (Zhu et al., 2020), as well as post-translational modifications *in vivo* (Bouclon et al., 2017). Besides odorant transport, other functionalities, such as scavenging, i.e. clearing ORs after molecular detection and preventing overstimulation and oversaturation, may also be relevant in the olfactory sense (Nakanishi et al., 2024). Figure 1 shows a simplified overview of the olfactory process at the olfactory cleft mucus (OCM), which represents the interface between air and the environment surrounding the olfactory neurons. OBPs are usually active in the monomeric form of the proteins. This is made possible through a highly conserved disulfide bond that fixates the C-terminal alpha helix to the wall of the beta-barrel. An exception is bovine OBP which lacks the necessary cysteine residues and is only stable in the form of a dimer (Ramoni et al., 2008,Vincent et al., 2004). It should also be noted that mammalian OBPs are structurally very different from insect OBPs who have also been reported to have distinct functions and mechanisms (Rihani et al., 2021; Abendroth et al., 2023). Selected examples of insect OBPs depicting their characteristic alpha helical structure are provided in the supporting information (see Figure S1). The stable hydrophobic beta-barrel structure of mammalian OBPs is a characteristic of the lipocalin protein superfamily to which they belong. Lipocalins are well studied and known to transport a variety of different hydrophobic molecules such as retinol, retinoic acid, aromatic compounds, and fatty acids (Breustedt et al., 2006). Altogether, 37 members of the lipocalin family have been identified in the human genome (Du et al., 2015). In humans, they are found in blood plasma, tears, genital secretion, and are mainly active in transporting hydrophobic molecules and binding to cell surface receptors (Charkoftaki et al., 2019). Selected human lipocalins, all showing the typical beta-barrel calix and the leaning alpha helix, and some of the corresponding ligands are depicted in Figure S2 and Figure S3, respectively. Table S1 provides an overview of the different members of the lipocalin family and their reported functions as listed in the Human Protein Atlas (HPA).

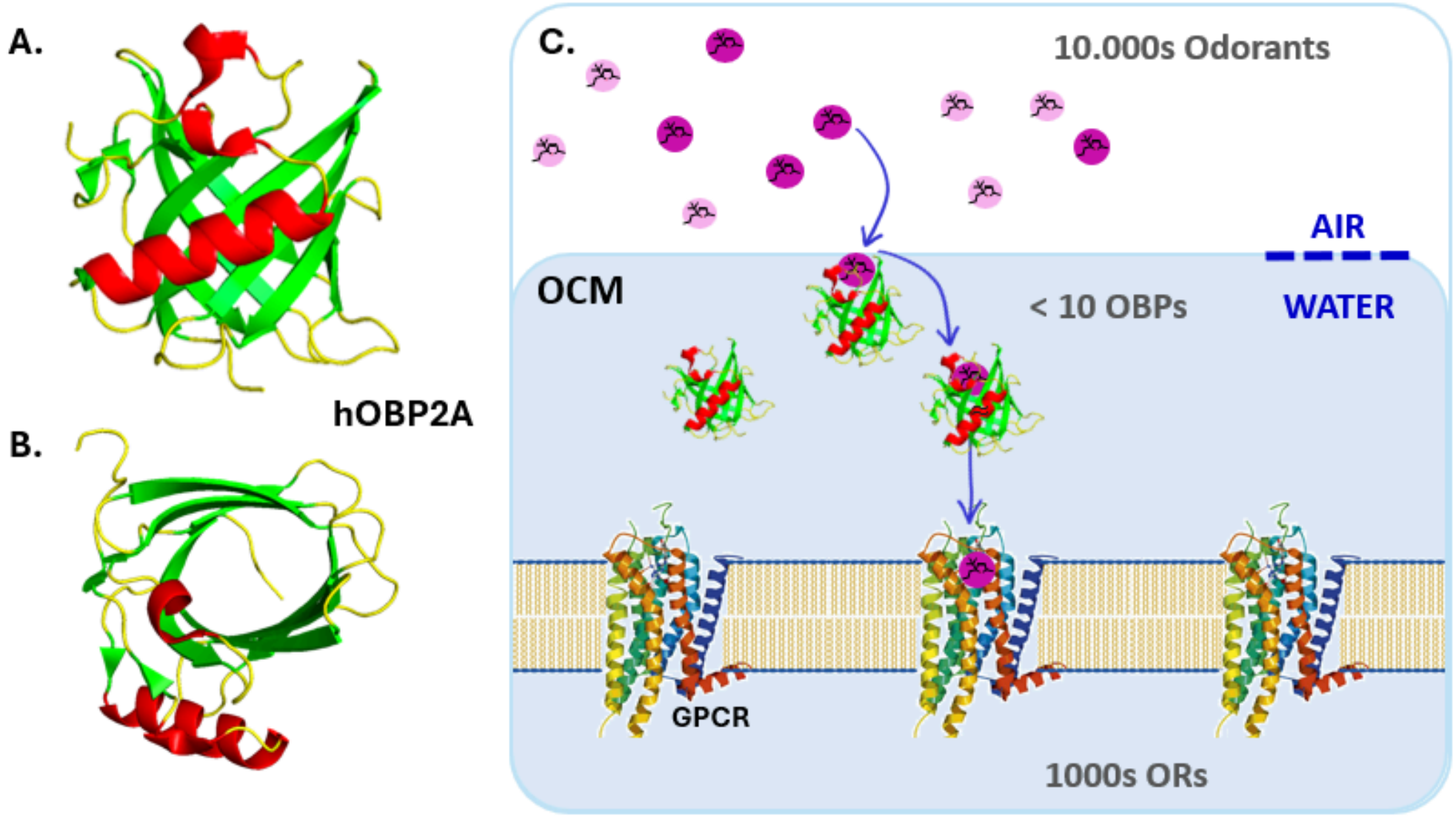

**Figure 1. Simplified overview of the function of mammalian OBPs in olfaction. A**: View on the alpha helix fixated on the beta-barrel structure of human OBP2A (hOBP2A, PDB-ID: 4RUN (Schiefner et al., 2015)). **B**. View along the hydrophobic calyx of hOBP2A (PDB-ID: 4RUN). C. Schematic overview of the transporter function of OBPs within the olfactory cleft mucus (OCM). Their main function is suggested to aid the transfer of hydrophobes (odorants) from the gas into the hydrophilic mucus towards the olfactory G protein coupled receptors (GPCR) embedded in the membrane. Note that while mammals usually possess less than 10 OBPs, up to 2000 olfactory receptors (ORs) may be present to detect the different odorants. Humans have approximately 400 different ORs (Orecchioni et al., 2022).

Among the many human lipocalins, at least three are implicated in cancer (Table S1). Lipocalin 2 (LCN2), also known as NGAL, has been the most extensively studied. In ovarian cancer, LCN2 was markedly overexpressed and associated with tumor differentiation, playing a role in iron transport and immune response modulation (Cho and Kim, 2009). In breast cancer, it induces epithelial-to-mesenchymal transition (EMT) by downregulating the estrogen receptor α (ERα), which increases the expression of Slug, a key EMT transcription factor, thereby promoting the mesenchymal phenotype and enhancing cell invasiveness and metastasis (Yang et al., 2009). In colorectal cancer, LCN2 is implicated in promoting tumor progression through the induction of the SRC/AKT/ERK signaling cascades (Zhang et al., 2021). These findings suggest that lipocalins may be involved in the tumor microenvironment and cancer cell physiology. However, the precise pathways through which lipocalins influence cancer remain to be fully elucidated. It should be noted that next to the impact of LCN2 in cancer, ORs have also been shown to regulate cancer cell activities, in addition to their core function of odor detection in olfaction (Chung et al., 2022).

So far, two OBPs have been reported in humans: hOBP2A and hOBP2B, that were initially identified based on their sequence identity (15.3% - 40.6%) to their rat OBP counterparts (OBPI and OBPII) (Pes et al., 1998; White et al., 2009). hOBP2A has been reported to be expressed in the nasal structures, salivary and lacrimal glands, and the lung, while hOBP2B is expressed in genital sphere organs such as the prostate and mammary glands (Lacazette, 2000). A deeper understanding of hOBPs emerged with the elucidation of the crystal structure of OBP2A and with the detection of OBP2A within samples of the olfactory cleft mucus (OCM) (Débat et al., 2007). While it has been shown that olfactory detection is possible in the absence of OBPs (Staiano et al., 2007, El Kazzy et al., 2025), their reported presence in the mucus covering the olfactory cleft, where the sensory olfactory epithelium is located, suggest their role as odorant carriers (Briand et al., 2002), which would in agreement with a function of “lowering” the barrier for odorant transfer from the gas to the solvated phase. To exert such a transporter function, OBPs need to be concentrated and operationable at the air-mucus interface. Several studies indicate that OBPs may facilitate the solubilization of hydrophobic odor molecules in the aqueous mucus layer, thus facilitating odorant transport through the hydrophilic mucus (Paesani et al., 2025; Hajjar et al., 2006). Nevertheless, labeling these newly discovered lipocalin members as odorant transporters may have restricted their study to putative functions in the olfactory sense, thus narrowing the possibility to identify and understand their full functionalities in other areas of the body. While the primary role of OBPs is traditionally associated with olfactory detection, emerging evidence suggests that these proteins also play significant roles in non-olfactory tissues (Ferrer et al., 2016; Sun et al., 2018). Experimental evidence of OBPs localization in nasal glands and secretions, emphasizes their secretory nature and potential involvement in protective and modulatory roles in epithelial barriers (Pevsner et al., 1986). OBPs show a functional diversity across organs, noting their expression in a variety of tissues, including reproductive and digestive organs, suggesting broader physiological roles (Sun et al., 2018). This could expand the relevance of OBPs from mere odorant carriers to multifunctional proteins essential in diverse physiological contexts beyond olfaction related tissues. Their expression in the reproductive organs, together with their ability to transport hydrophobic molecules such as hormones is particularly intriguing as several cancer types are hormone dependent. OBPs could easily function as hormone transporters, contributing to tumorigenesis or other diseases such as Parkinson’s disease (Melis et al., 2019). For the latter, recent studies identified the tendency of bOBP to form larger protein aggregates under different experimental conditions (Stepanenko et al., 2023). The work highlights the potential usefulness of using OBPs to study amyloid fibril formation, and as model systems to better understand and mimic the oligomerisation mechanisms underlying neurodegenerative diseases (Stepanenko et al., 2024a; Sulatskaya et al., 2024). However, in this structured review, we focus on investigating the potential involvement of OBPs in cancer, thus opening additional new avenues to understand and exploit the function of these proteins. This hypothesis is further supported by the diverse roles of lipocalins such as LCN2 in modulating immune responses and cancers (Rodvold et al., 2012), and isolated reports of potential hOBP2A involvement in prostate cancer (Jeong et al., 2023) and colorectal cancer (Cervena et al., 2021). In this review, we address the question whether OBPs may have a broader impact and more sophisticated functions than anticipated so far. We focus on the putative functions of hOBP2A and hOBP2B outside of the olfactory sense, exploring their implications in cancer biology and the broader physiological roles they may play in human health. Figure 2 provides an overview of our approach, which leverages available information from the literature and various databases, such as the HPA (Uhlén et al., 2015, 2019; Uhlen et al., 2019) and the cancer genome atlas (TCGA) (Tomczak et al., 2015), to carry out a data-driven study to elucidate the underlying function of OBPs under normal and cancerous conditions. In addition, insights from structural bioinformatics allow us to analyse the potential impact of mutations on the functions of these two proteins. Finally, we conclude our overview with a discussion of future directions and the most relevant applications OBPs and other members of the lipocalin family, as provided by current literature.

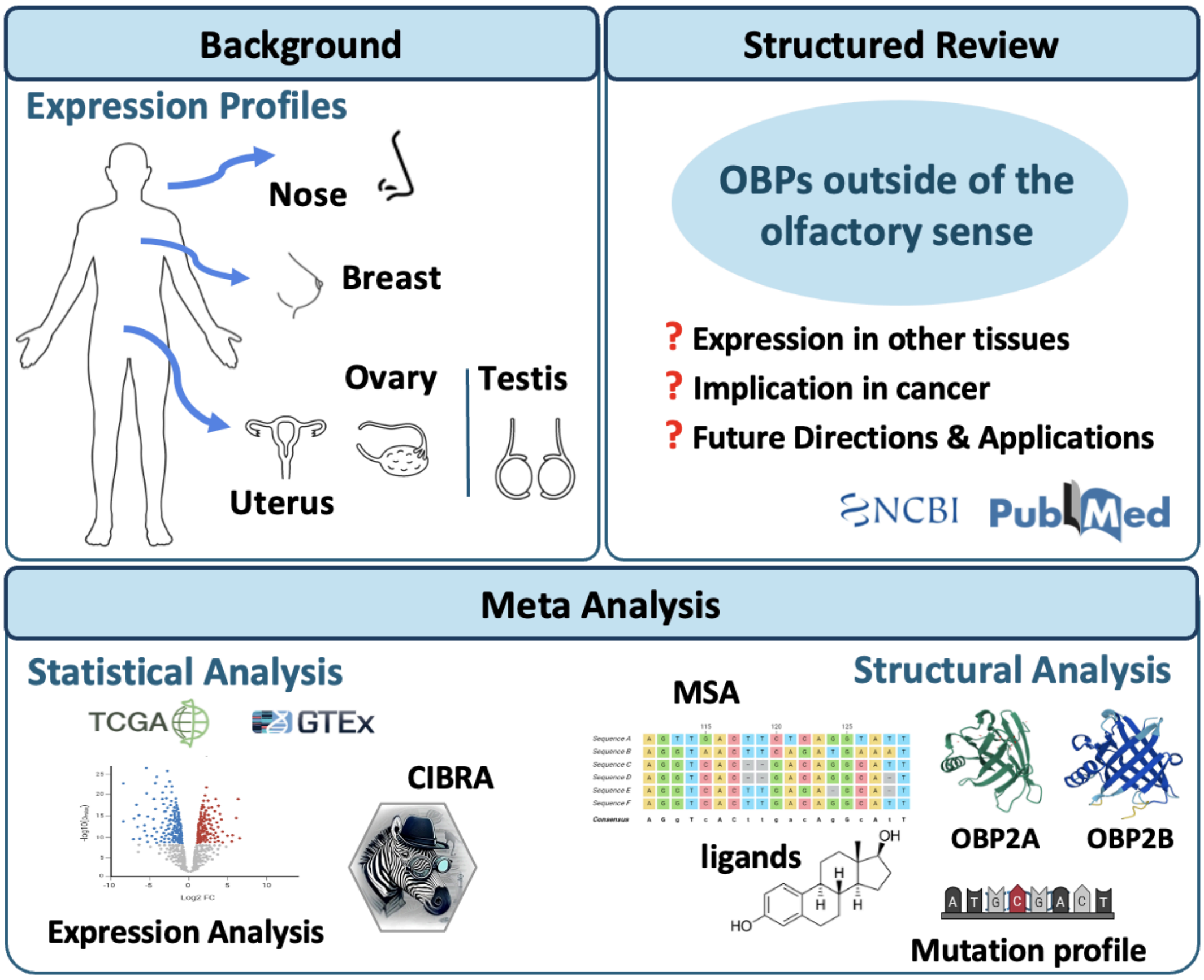

**Figure 2**. **Schematic figure highlighting the approach and research question underlying this work.** Due to the reported expression of human OBPs in tissues beyond the olfactory tract, in particular the reproductive organs, we investigate the potential role of OBPs outside the olfactory sense through a systematic meta-analysis of existing data (cf. methods section). Using differential gene expression (DGE) analysis on available data from the cancer genome atlas (TCGA), multiple sequence alignment (MSA), and protein function analysis, next to a detailed literature review, we look into their potential involvement in hormone transport and cancer tumorigenesis.

## Occurrence of OBPs in tissues outside of the olfactory sense

### OBP2A and OBP2B are expressed in tissues outside the olfactory sense and hormone related cancer types

To see whether OBPs are expressed outside of the olfactory sense, the HPA database was used (Uhlén et al., 2015; Thul et al., 2017; Sjöstedt et al., 2020), which provides an overview of reported RNA and protein expression levels throughout the body. For OBP2A and OBP2B, the RNA-expression data from the HPA showed that both proteins are noticeably expressed in healthy tissues unrelated to olfaction. Figure 3 shows an overview of the RNA expression levels in different tissues, including male and female tissues. It can be seen that both OBP genes show high RNA expression in the male and female reproductive tissues. While OBP2A is expressed at a high level in female tissues (e.g., the fallopian tubes), OBP2B is significantly expressed in male tissues (testis). Since both OBP2A and OBP2B are predominantly associated with the olfactory system, this high expression in the female and male tissues raises the intriguing question of whether they contribute to the function or pathology of these non-olfactory tissues. Additionally, as OBPs (and lipocalins in general) are known to bind hydrophobic molecules, it is noticeable that they are both concentrated in tissues whose functions are regulated by hormones. Given that the hormones of the reproductive systems, i.e., estrogen, progesterone, testosterone, are hydrophobic sterane derivatives similar to odorants (Oren et al., 2004), OBPs are therefore ideal candidates to function as hormone transporters. Based on this observation, we hypothesize that OBPs may play a role in hormone transport and/or hormone regulation in these tissues.

For a more comprehensive investigation, we dive into protein expression levels of OBP2A and OBP2B in different reproductive tissues as reported in the Protein Abundance database (PaxDB) (Wang et al., 2015; Huang et al., 2023) and other proteomics repositories (Lonsdale et al., 2013; Thangudu et al., 2024). Interestingly, the reported protein expression levels concentrated in the male and female reproductive tissues are higher than the expression levels in the olfactory system. This may indicate that only low expression levels of OBPs are required for functionality in the olfactory sense and potentially support the hypothesis that odorant transport through the mucus may also be achieved spontaneously with OBPs, however, most likely at a slower rate.

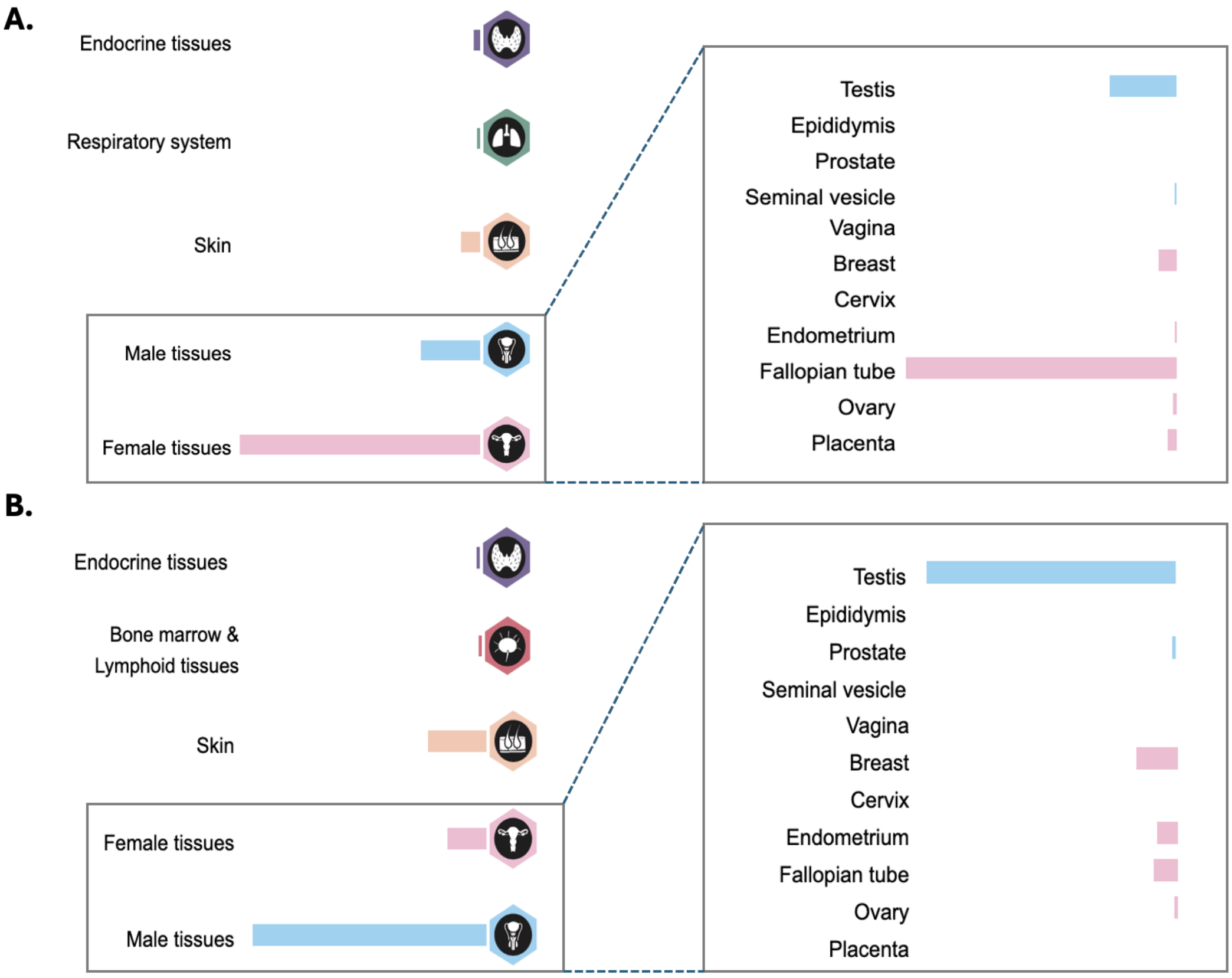

**Figure 3. OBPs RNA Expression Levels in Male and Female Healthy Selected Tissues.** The HPA (Uhlén et al., 2015; Sjöstedt et al., 2020) provides RNA expression summaries for A) OBP2A and B) OBP2B in various human tissues. Expression levels are reported in normalized expression units, measured as Transcripts Per Million (nTPM), derived from both HPA and Genotype-Tissue Expression (GTEx) RNA-seq data (Figure taken from https://www.proteinatlas.org/ - 23/06/2025). The highest enrichment for OBP2a and OBP2B is found in the fallopian tubes and testis, respectively. Detailed information on these expression levels is available in the tissue atlas for each protein (Lonsdale et al., 2013).

In addition to expression levels in healthy tissues, the HPA further reports the RNA and protein expression of OBPs in different cancer tissues, such as breast and ovary (see supporting information Figure S4 for details). In addition, the protein expression of OBP2A and OBP2B in cancer cases on the Proteomic Data Commons (PDC) portal (Thangudu et al., 2024) shows that OBP2A is differentially expressed in ovarian cancer and OBP2B in breast cancer. Since it is known that hormones of the reproductive system play a pivotal role in the development and progression of certain cancers, such as breast (Yang et al., 2009), ovarian (Cho and Kim, 2009), and prostate cancers (Jeong et al., 2023). If OBPs indeed have a hormone-transporting function in reproductive tissues, it is worthwhile investigating whether they may also be implicated in the molecular mechanisms underlying hormone-driven cancers. An overview of the RNA and protein expression of OBP2A and OBP2B reported for both, normal and cancer tissues, in different databases is provided in Table 1. Due to their characteristics as small secreted proteins that can transport hydrophobic molecules, they can play roles in lots of physiological activities such as immunity, metabolism, and cellular signaling. Olfactory receptors are included in this overview to demonstrate that at least two receptors, OR51E1 and OR51E2, are also expressed in non-olfactory tissues such as normal reproductive organs (see Figure S5) and cancer tissues (Li et al., 2021; Chung et al., 2022; Orecchioni et al., 2022). For a complete overview of RNA and protein expression patterns of human lipocalins and ORs across tissues and an overview of known mechanisms of lipocalins in cancer see supporting materials (see Table S1, Figure S6). In the case of human lipocalins, almost every one of the family members displays RNA expression in cancer, with the exception of LCN9, while their protein expression in cancer tissues is only reported for half of them. Interestingly, the reported protein expressions of lipocalins are mainly related to reproductive cancers and several lipocalins have been directly linked to tumorigenesis and metastasis (Bratt, 2000; Du et al., 2015), e.g., LCN1 and LCN2 in breast (Yang et al., 2009) and uterine cancer (Mannelqvist et al., 2012) (Table S1 in the supporting information). The ectopic expression of ORs is shown across multiple cancer types, including breast, prostate, and lung cancers (Chung et al., 2022). The functional roles of ORs appearing in reproductive tumor microenvironments can be a potential connection with OBPs influence in those cancers. Furthermore, the pro-metastatic role of OR5B21 in breast cancer reveals that ORs promote cancer invasion and metastasis through epithelial-mesenchymal transition (EMT) (Li et al., 2021). This may hint  that ORs play a role in aggressive tumors. Expanding on this, the overexpression of OR2T6 has been reported to  the invasive capabilities of cancer cells (Li et al., 2019). The above evidence provides an idea that ORs can act beyond sensory perception and contribute to oncogenesis. It should be noted that ORs (see also Introduction section), which are the key to olfactory detection mechanisms, have been shown to play a role in cancer progression (Chung et al., 2022). It is therefore possible that OBPs are simply co-expressed as a side-effect of ORs or lipocalins. Another interesting aspect about the function of OBPs may be if they are able to bind specifically to ORs or membranes. For insect OBPs it has been shown that they are able to bind ORs (Leal, 2013), while this is not explicitly the case for mammal OBPs (Pelosi, 1998). At a higher level, lipocalins are also known to interact with membrane receptors, e.g., megalin (LRP2) is a broad receptor of lipocalins and LCN1 and LCN2 have known binding site for molecular uptake  (Wojnar et al., 2003; Cabedo Martinez et al., 2016). Although binding to an OR may not be necessary for OBPs within their function in the olfactory sense, it cannot be ruled out that interactions within other membrane proteins or membranes in general plays a role in other tissues or cancerous environments.

**Table 1.** RNA and protein expression of OBP2A and OBP2B. Reported cases in both normal and cancer tissues across 4 different databases. The highest expression levels are highlighted in bold.

| Conditions | Database | OBP2A | OBP2B |
|---|---|---|---|
| RNA - normal | GTEx[1] | **Fallopian tube**, Testis, Skin, Breast, Ovarian | **Testis**, Breast, Skin, Fallopian tube |
| | HPA[2,3] | **Fallopian tube**, Breast, Testis, Skin | **Testis**, Breast, Fallopian tube, Endometrium, Skin |
| RNA - cancer | HPA | **Ovarian**, Breast | **Breast**, Ovarian |
| Protein - normal | PaxDb[4] | **Testis** (13.9), Gut (11), cerebrospinal fluid (6.63), Fallopian tube (0.09)[6] | **Brain** (10), Placenta (7.31), Fallopian tube (2.36), Testis (1.87), esophagus (1.82), Endometrium (0.24) Saliva secreting gland (0.2)[7] |
| Protein - cancer | PDC[5] | **Ovarian** | **Breast** |

[1]GTEx: Genotype-Tissue Expression Portal (Lonsdale et al., 2013).

[2]HPA: The Human Protein Atlas (Uhlén et al., 2015, 2019; Uhlen et al., 2019).

[3] Note that the expression of OBP2A and OBP2B are higher in female and male tissues, respectively (cf. Fig. 3).

[4]PaxDB: Protein Abundance Database (Wang et al., 2015; Huang et al., 2023).

[5]PDC - Proteomic Data Commons. PDC was supported by National Cancer Institute (NCI) (Thangudu et al., 2024).

[6,7]Concentration levels provided in descending order as reported in PaxDB.

**OBP2A and OBP2B share a high sequence similarity**

As protein structure is directly linked to protein function, we have looked into the experimental crystal structure of OBP2A (PDB-ID: 4RUN (Schiefner et al., 2015)) and the AlphaFold3 (Jumper et al., 2021) predicted structure of OBP2B along with their sequence and corresponding point mutations to further elucidate their potential functions. The two proteins exhibit a sequence identity of 90% with a total of 17 different residues. The corresponding MSA shown in Figure 4b illustrates the high similarity of the two sequences. Figure 4a highlights the positions of the amino acids that differ from OBP2B. These 17 non-conserved residues are spread over the entire protein chain, including the alpha-helix and the beta-barrel, which does not suggest different functionalities between the two proteins (see Table S2 and Figure S7-8 for details). Rather, with the high sequence similarity, the structure of OBP2B can be expected to be close to identical to OBP2A, and both proteins will probably be able to transport a large amount of similar hydrophobic molecules. The conserved hydrophobic residues throughout the structures are crucial to maintain the typical beta-barrel structure of the transporter protein. Both proteins are therefore also likely to perform similar functions. However, some of the differences, in particular the 5 residues that shift from hydrophilic to hydrophobic physicochemical properties that are partially located close to the binding pocket, are likely to impact the affinity or binding properties of proteins towards different ligands or to yield different roles in a given tissue type (See Figure S8). Changes may include tissue-specific expression or regulatory properties. Despite the high similarity at the protein level, they may be differentially regulated by different promoters or tissue specific signals. Using porcine OBP as a case study, previous works showed that selected mutations within the binding pocket can even lead to chiral discrimination between the enantiomers of carvone (Paesani et al., 2025). In order to get a full understanding of hOBPs and their relationship to other members of the lipocalin family, as well as ORs, we verified the chromosomal location reported in genecards (Safran et al., 2021). Conserving sites of the coded gene of odorant-binding proteins, the *OBP2A* and *OBP2B* genes are located close by on chromosome 9, with *OBP2A* and *OBP2B* on Chr 9q34.2 and Chr 9q34.3, respectively (Safran et al., 2021). Figure 4c shows the distribution of the different lipocalins on chromosome 9. Interestingly, most kernel lipocalins are in the same position as *OBP2A* in Chr 9q34.3. *LCN2* exhibits a unique locus in Chr 9q34.11. In contrast, ORs are located on chromosome 11.

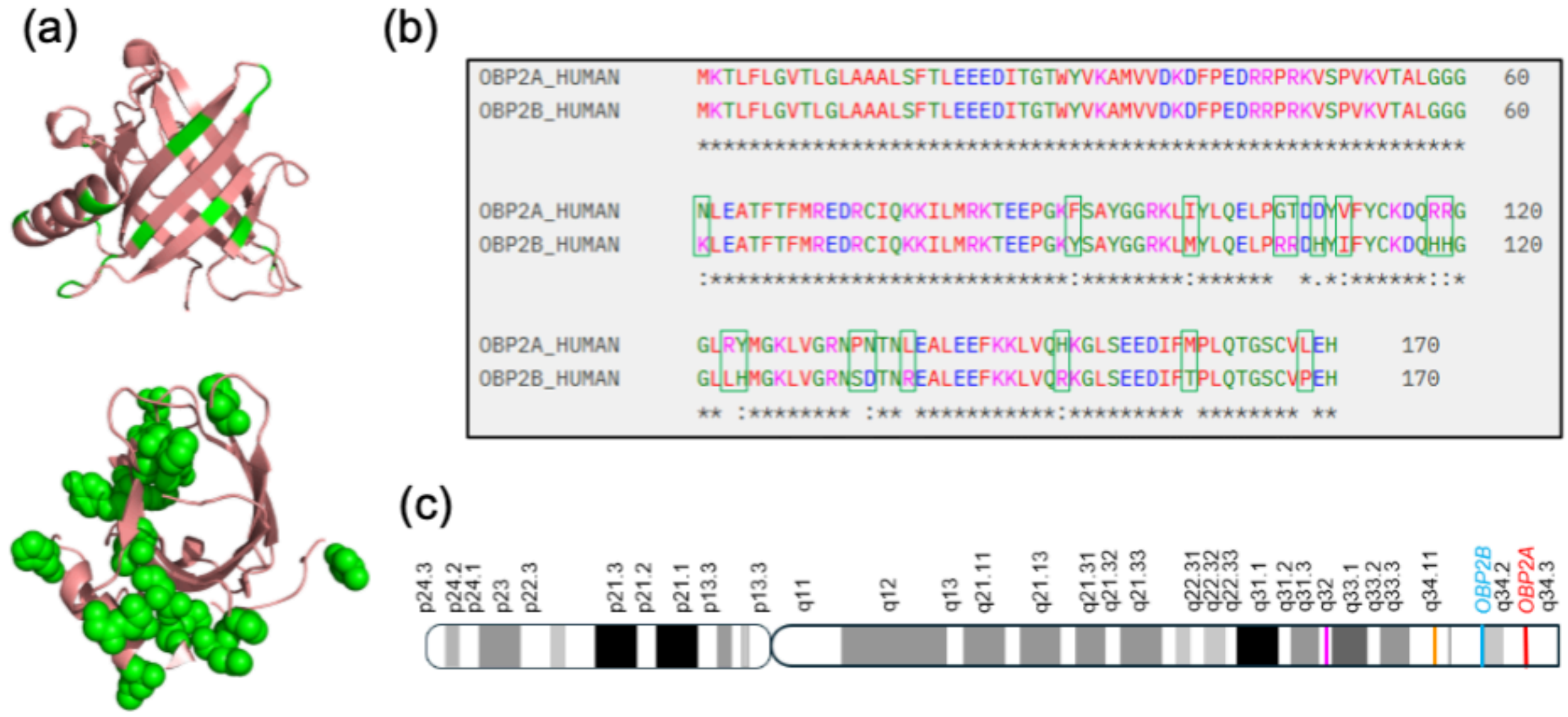

**Figure 4. Structure, Sequence, and Chromosomal Location of hOBPs.** (a) Typical 3-Dimensional crystal structure of hOBP2A (PDB-ID: 4RUN). The different sequences are visualized in green for both side view with cartoons (top) and top view with spheres (bottom) in Pymol. Corresponding green residues represent sequence differences implied in (b). The main chain (alpha-helix and beta-barrel) color of hOBP2A is salmon. (b) MSA of hOBP2A and hOBP2B protein sequences. Sequence differences are demonstrated in green rectangles. hOBP2A and hOBP2B share 90% identity in sequence similarity according to NCBI Blast. Sequence alignment representation: '*' indicates identical residues, ':' denotes residues with similar properties, '.' represents residues with slight similarity, and a space signifies completely different residues. (c) Chromosomal location of OBP2A (highlighted in blue) and OBP2B (highlighted in red) both found on chromosome 9, along with other human lipocalin superfamily members found on location 9q34.2, 9q34.3, 9q34.11, and 9q34.32 (figure adapted from Genecards (Safran et al., 2021)). Genomic localization of lipocalin family genes on chromosome 9: Genes at 9q32 (magenta) include *AMBP, ORM1,* and *ORM2*; at 9q34.11 (orange), *LCN2*; at 9q34.2 (blue), *OBP2B*; and at 9q34.3 (red), *OBP2A, LCN1, LCN6, LCN8, LCN9, LCN10, LCN12, LCN15, LCNL1,* and *PTGDS*. Note that olfactory receptors are located on chromosome 11.

**Network and co-expression patterns of OBP2A and OBP2B**

Similar to mammalian OBPs, insect OBPs have been reported to be expressed in specific tissues including antennae as well as reproductive and gustatory tissues (Hu et al., 2016). However, they have been shown to be co-expressed with olfactory receptors and to cluster on co-expression, experiments and corresponding protein homology with genes involved in signal transduction (Mika and Benton, 2021; Task et al., 2022), detoxification and neural signaling, which all strongly suggests their implication in olfactory pathways (see reported network in the STRING database (Huang et al., 2023) - data not shown). In mammalian, and particularly human olfaction, this is not as straightforward. Although independent sources have identified OBP2A and OBP2B to be expressed in the nasal mucus (Verbeurgt et al., 2014), there are also studies reporting that unlike mouse OBPs, human OBP2A and OBP2B do not show signs of enhanced expression in sensory tissue (Olender et al., 2016). Figure 5 visualizes the known protein connection network between OBP2A and OBP2B as reported in the STRING database (Huang et al., 2023). There, OBPs are indeed only reported to be co-expressed with other lipocalin family members, most likely due to their location on chromosome 9 (see also previous section), but there is no reported co-expression with other proteins from the olfactory system. The lack of identified gene expression networks and pathways for human OBPs makes it difficult to straightforwardly associate their role in olfactory pathways. While their expression is most likely directly linked to lipocalin expression, this does not necessarily exclude the putative implication of OBP2A and OBP2B in cancer progression or other functions across the body. Note that the reported links are not solely based on text mining but that several co-expression patterns are reported between OBP2A and OBP2B, as well as to other members of the family, namely LCN8 and LCN12, which are both predominantly reported in the male reproductive tissues in the HPA. It should also be noted that the STRING database does not report any link to LCN1 and LCN2 in the network. Altogether, it can be said that the role of OBPs in human olfaction are still disputed, these proteins are rather intriguing and their initial name may indeed have been somewhat hasty and limiting to fully study and exploit the functionality within the context of olfaction and beyond. Within the notable expression profiles of OBPs in reproductive tissues and some cancer types, fully understanding them will require to go beyond their expression profiles, and to look closer into their mutation profiles and the structural and functional implication that these may involve.

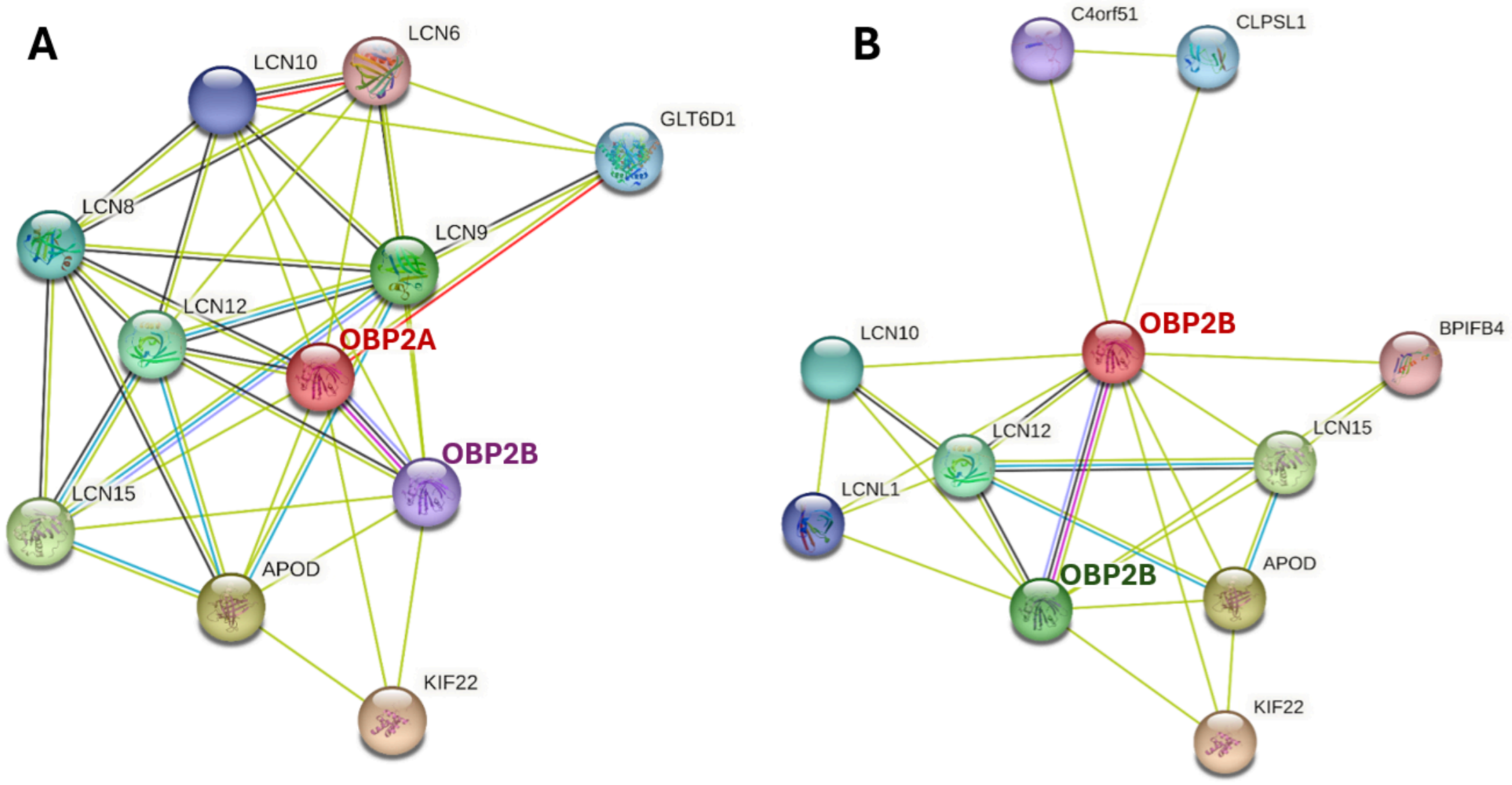

**Figure 5. PaxDb STRING protein interaction network of OBP2A (left) and OBP2B (right).** Color code: Colored nodes, query proteins and first shell of interactions (refer to amino acids or atoms that directly interact with the core such as a ligand or active site); White nodes, second shell of interactions (not directly contact the core of molecule but interact with residues involved in first shell interactions). For node content: Empty nodes, proteins of unknown 3D structure; Filled nodes, some 3D structure is known or predicted. Classification of gene/protein interactions: Known interactions are indicated in light blue (curated databases) and magenta (experimentally determined). Predicted interactions include green (gene neighborhood), red (gene fusions), and dark blue (gene co-occurrence). Additional associations are represented by light green (text mining), black (co-expression), and light purple (protein homology).

**The genomic landscape of different cancer types indicates alterations in OBP genes**

Using the available data from the HPA, we performed differential gene expression analysis to determine whether their expression is significantly altered in different cancer types. Specifically, DGE was conducted to identify the expression patterns of OBP2A, OBP2B and other lipocalins, as well as known cancer genes (cf. Table S3 for the list of included known cancer biomarkers) in six different cancer types: ovarian, breast, uterine, prostate, colorectal, and lung cancers (TCGA database(Tomczak et al., 2015)) and normal (GTEx database (Lonsdale et al., 2013)). The datasets used were screened out based on the filter parameters shown in Table S4. The results of the DGE for ovarian cancer are depicted in the volcano plots shown in Figure 6 (see Figure S9 for the remaining breast, uterine, prostate, lung, melanoma plots). The analysis reveals that both OBPs are upregulated in ovarian and breast cancer, while OBP2B is upregulated only in ovarian cancer.

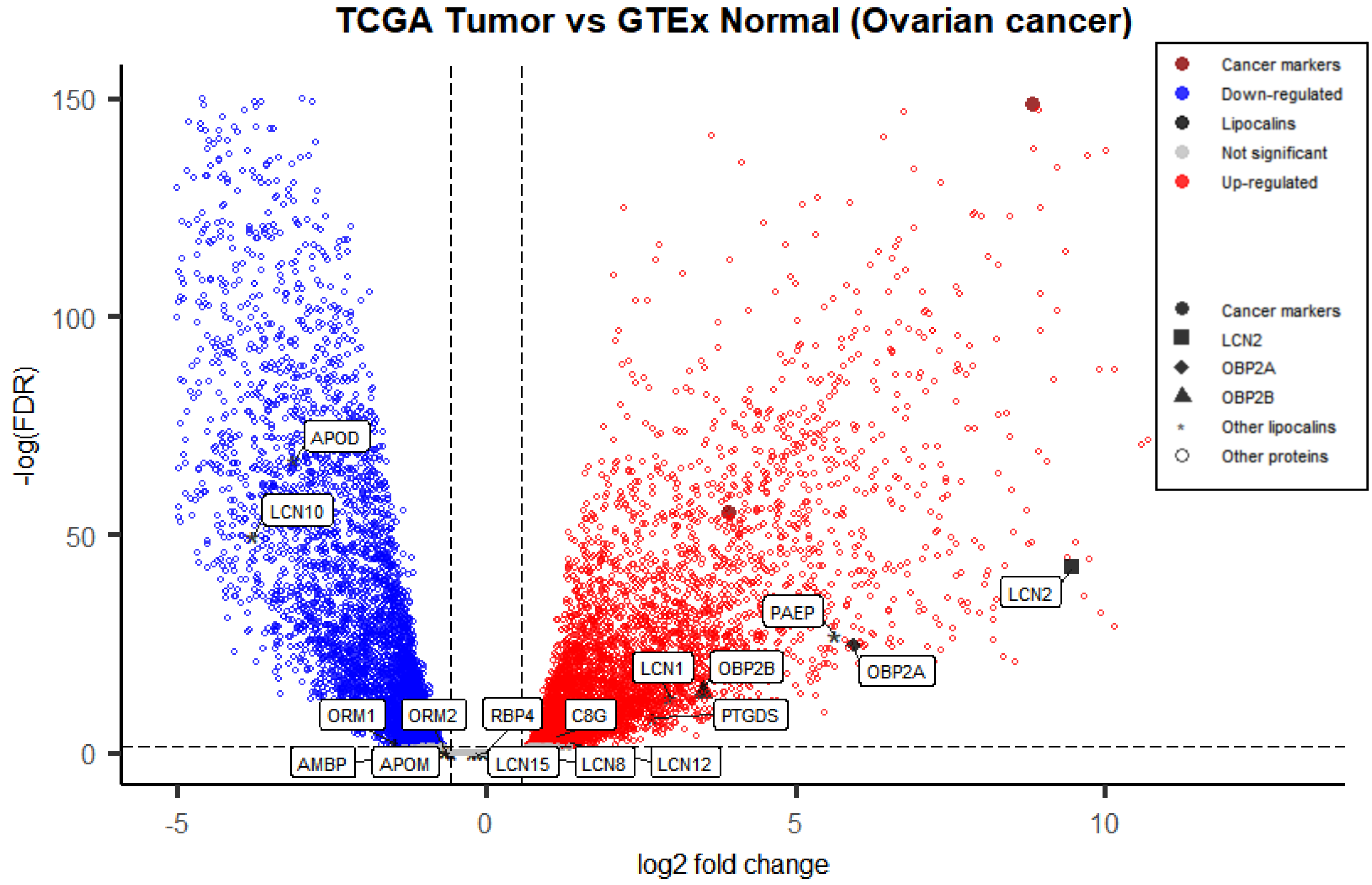

**Figure 6. Volcano plots visualizing the differential gene expression analysis of OBP2A, OBP2B and lipocalins in ovarian cancer.** Volcano plots illustrate the differential expression profiles between normal and tumor samples in ovarian cancer to display the distribution of OBP2A and OBP2B based on their statistical significance and magnitude of change. The plot displays log2 fold change on the x-axis and –log10 adjusted p-value on the y-axis. Genes are color-coded: blue indicates downregulation (adjusted p-value $< 0.05$ and log2 fold change $< -1$) , and red represents upregulation (adjusted p-value $< 0.05$ and log2 fold change $> 1$). Non-significant genes are represented in gray. In this analysis, human lipocalins and cancer markers (as detailed in Table S3) are labeled in the plots.

Given the strong link between genetic mutations and cancer, we also examined the mutational profiles of OBP2A, OBP2B, and other human lipocalins using data from TCGA. From the oncoprint shown in Figure 7, it can be seen that the FABP family is most frequently mutated, with a mutation frequency of 53% among the human lipocalins across the cancers in the TCGA. APOD ranked in second place with a mutation rate of 48%. Most of the lipocalins were affected in at least 40% of patients. The lipocalins were mostly affected by amplifications and deletions that can span several genes. There is a high co-occurence of these mutations across the lipocalins that is potentially due to the close proximity of these genes on chromosome 9.

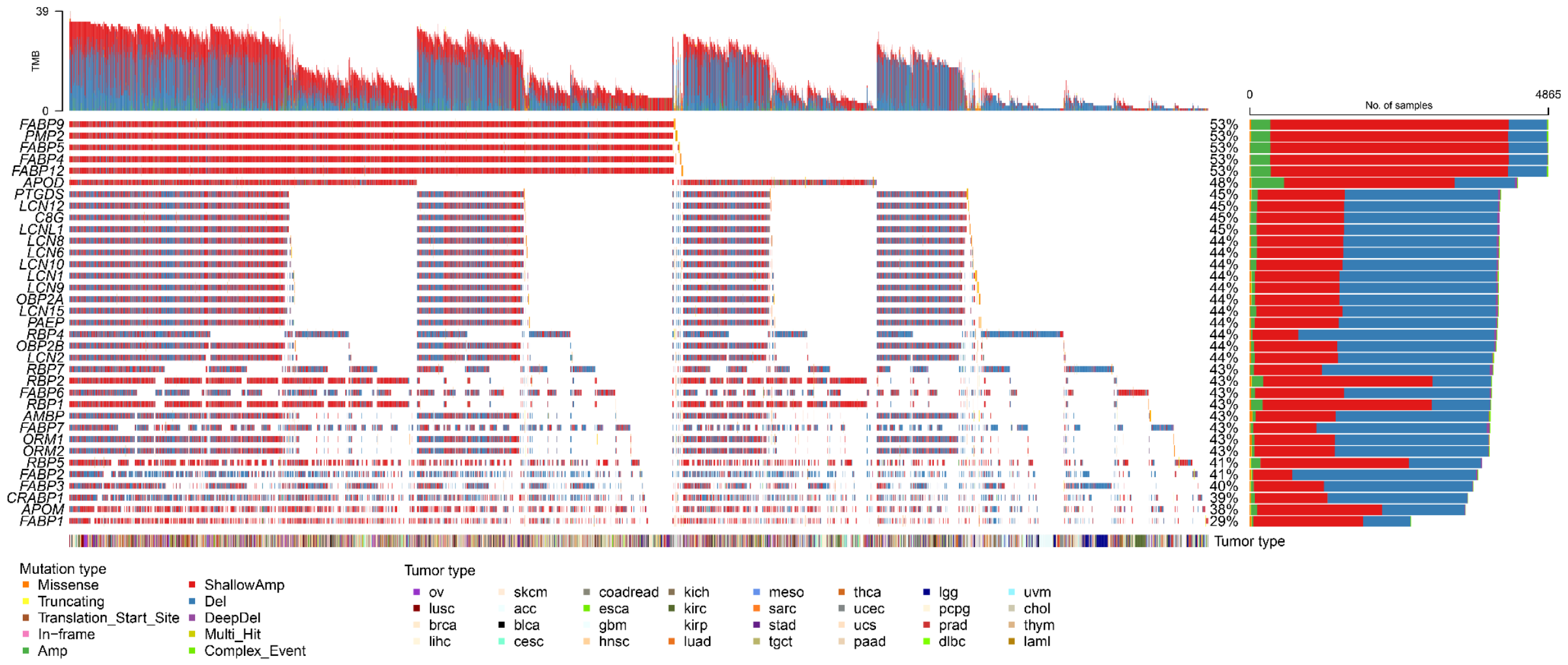

**Figure 7. Oncoplot of mutations in human lipocalins from TCGA.** Mutations are visualized for each lipocalin gene (rows) per patient (column) clustered on co-occurence. Mutations are categorized in missense, truncating, translation start size, in-frame, amplifications, shallow amplifications, deletions, deep deletions, multi hit and complex events. Tumor mutational burden (TMB) is color-coded at the top, ranging from 0 to 39 mutations per megabase (m/Mb). The percentage of profiled samples, and number of samples altered for each gene are displayed on the right.The tumor type is color coded on the bottom.

## Statistical analysis of the mutational landscape in different cancer types shows that CNAs represent the only significant genomic alteration, while the reported SNVs are unlikely to affect protein function.

The DGE analysis did not confirm an overexpression of OBP2A in prostate cancer, as observed in prostate cancer tissues that had shrunk during remission under androgen deprivation therapy (Jeong et al., 2023). However, the study suggests that OBP2A overexpression may be specifically induced by an androgen-deprivation state (Jeong et al., 2023), implying that the overexpression of OBP2A might require a very specific cancer microenvironment. This is in agreement with other studies in mammals that indicate the expression of OBPs may be influenced directly by external factors and developmental conditions (Todrank et al., 2011; Kuntová et al., 2018; Gonçalves et al., 2021). As for lipocalins among the six cancer types (ovarian, breast, uterine, prostate, lung and colorectal) analyzed in this study, LCN2 serves as an active factor in oncogenesis, and APOD was found to be downregulated in four of the selected cancer types (Figure S7). To assess if the differential expression of OBPs is due to mutations (CNAs and SNVs) affecting them, we have compared the expression of OBPs between patients with and without these mutations.

OBP2A and OBP2B show increased expression when gained and decreased expression when lost in ovarian cancer (Figure S10). Breast and uterine cancer show less pronounced differences. SNVs showed no change in expression. To further assess if the mutations within OBP2A and OBP2B have an impact, we have utilized the computational method CIBRA to measure the system-wide impact of these mutations (including gains and losses). From the CIBRA analysis, CNAs in OBP2A and OBP2B are associated with significant changes in the system-wide expression profiles of ovarian, breast and uterine cancer (Figure 8), the remaining figures for breast and uterine cancer see Figure S11.

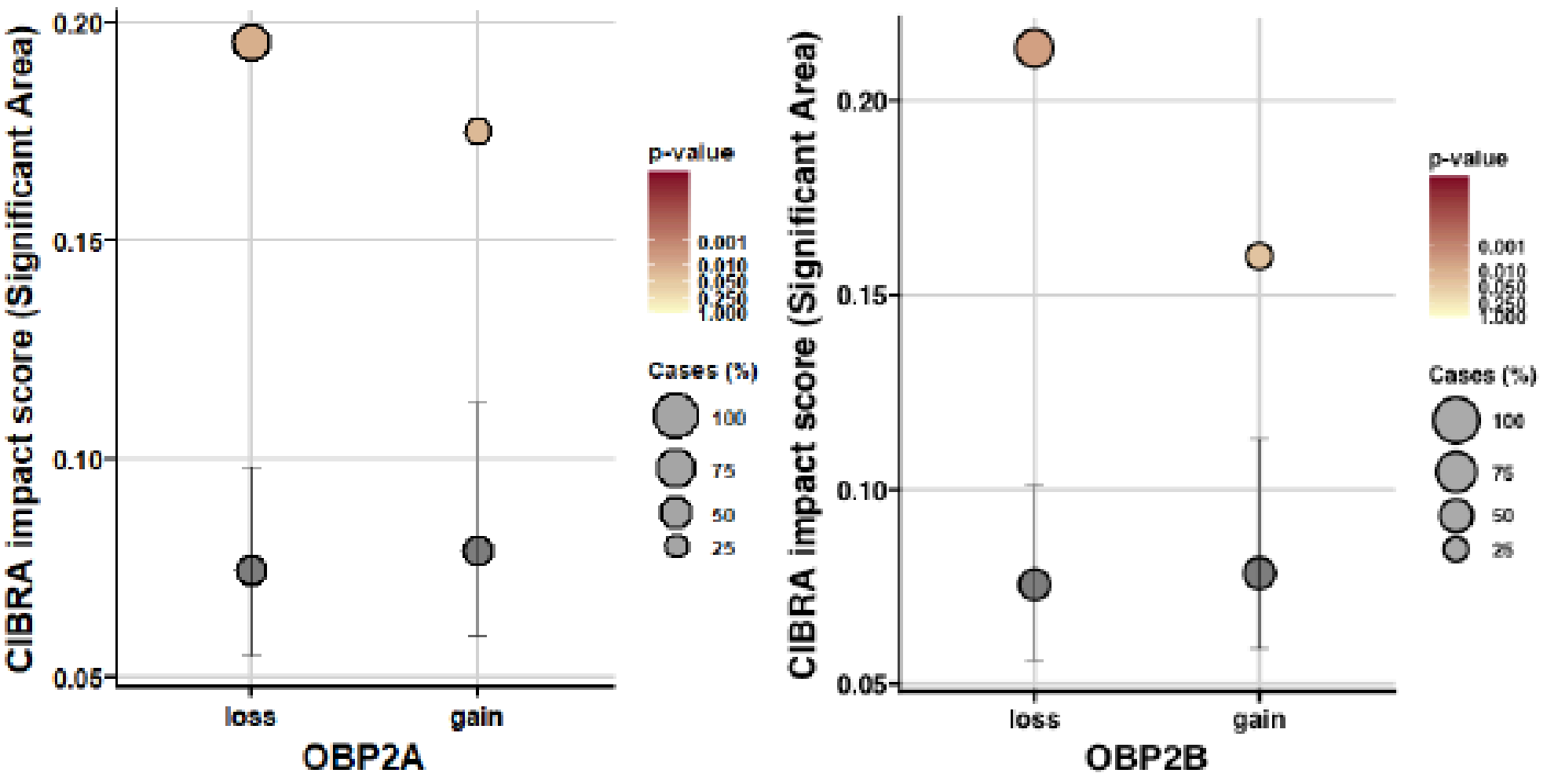

**Figure 8. CIBRA impact scores for OBP2A and OBP2B copy number alterations (CNA) in ovarian cancer.** Spheres represent the significant area of impact scores associated with either copy number loss or gain, as analyzed by the CIBRA pipeline. The size of the bubbles indicates the percentage of cases (%) affected by each alteration, and the color gradient corresponds to the p-value, where darker red shades denote stronger statistical significance (lower p-values); Light red suggest less significant results (higher p-values). The vertical axis represents the CIBRA impact score (significant area), which reflects the extent to which each CNA (loss or gain) influences the molecular profile in these cancers; higher values suggest a greater impact of the genetic alteration. For X-axis, loss: tumors where the gene is deleted (reduced copy number); gain: tumors where the gene is amplified (increase copy number). Error bars represent the interquartile range (IQR) of the data.

While the CIBRA results showed that copy number changes can affect the expression level of OBP2A and OBP2B, and SNVs did not result in expression changes. The mutations could still affect the protein structure and thereby the function.

To determine whether patient-derived SNVs may have an influence on the protein function, the physical-chemical properties of the corresponding mutated amino acid residues were assessed. Of all 44 mutations that were detected in OBP2A, 10 residues changed from hydrophilic to hydrophobic, and 3 changed the other way around. For OBP2B, 31 mutations were observed (see Figure S12). Among them, 4 changed properties from hydrophilic to hydrophobic, 2 from hydrophobic to hydrophilic. Most of them belong to missense mutations. The location of those mutated residues are highlighted in structures respectively in OBP2A (PDB-ID: 4RUN) and OBP2B (AlphaFold3 predicted structure) are depicted in Figure 9. To calculate the expected number of mutations due to random chance, we first determined the average mutation counts from the cancer type exhibiting the highest frequency of mutations to verify that the frequency of mutations is higher than random chance and meaningful (see Table S5 for further information).

Altogether, this does not directly imply a drastic change of functionality, but it is possible that at least the 6 mutations that include hydrophobic residues also impact the uptake capability and alter the binding profile of the proteins. Previous experimental works showed that mutated porcine OBP exhibits higher binding specificity towards carvone and may even be used to distinguish between different enantiomers of the molecules (Mulla et al., 2015). It may also be interesting to explore whether the reported mutation can impact the ability of OBPs to bind membrane receptors or ORs, a topic on which little is known so far. To fully understand the impact of these mutations on the structure and dynamic of the proteins, however, molecular dynamics simulations combined with guided bio-chemistry experiments to validate simulations may provide meaningful insight (Yi et al., 2015).

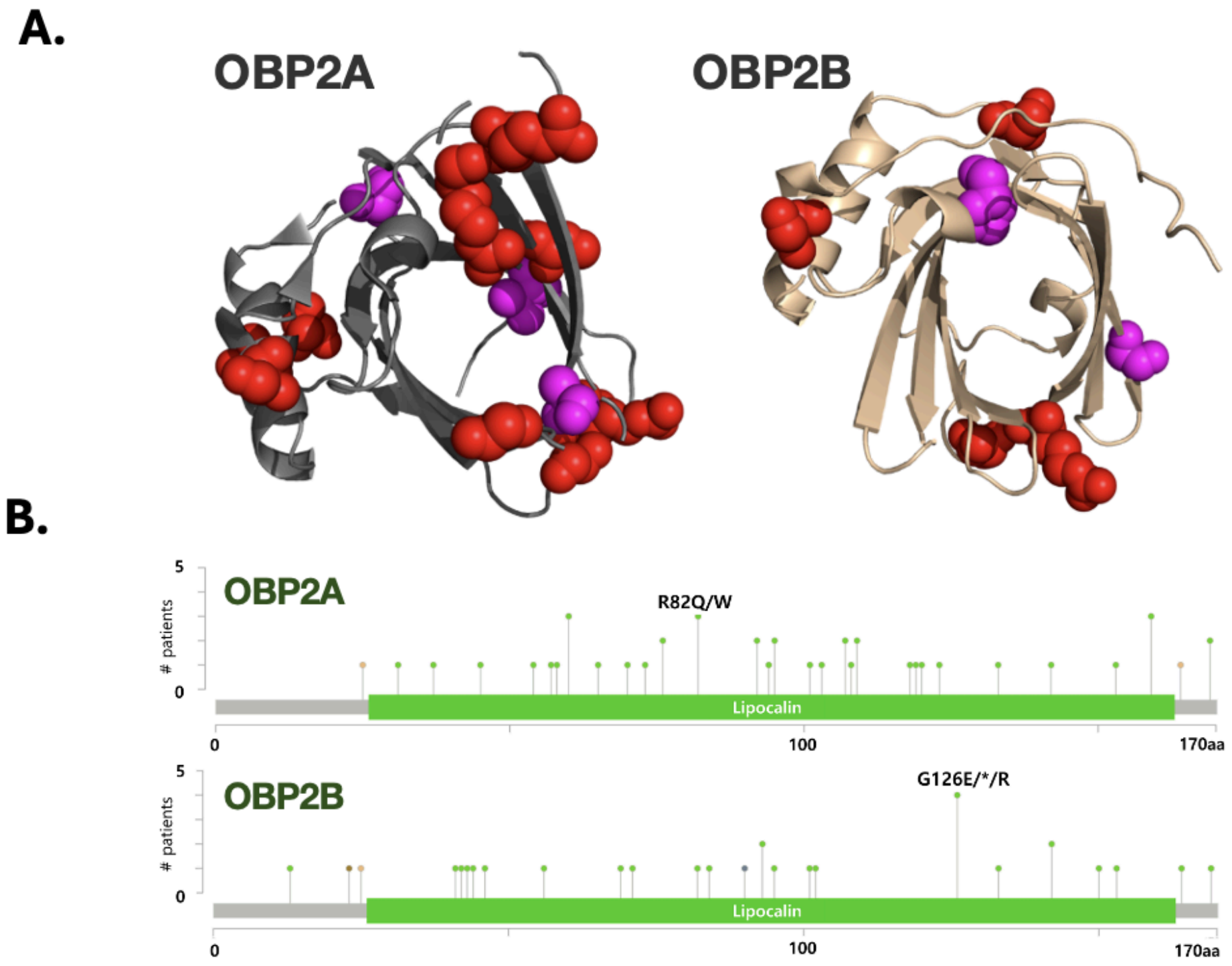

**Figure 9. Structural Visualization of OBP2A and OBP2B Mutations.** The figure depicts (A) mutated residues in OBP2A (left) and OBP2B (right) sampled across various cancers; (B) point mutations of OBP2A (up) and OBP2B (below) reported in cancer samples retrieved from cBioPortal are indicated at their respective positions with the number of patient samples with mutations at each amino acid position represented on the y-axis (specifically, one sample is one patient). Each dot represents a mutation, with the height showing how often it occurs. The most frequent missense mutations are labeled for each gene (green dots). The x-axis represents the amino acid positions from the start (0) to the end (170) of the protein. The Lipocalin domain spans most of the protein length indicating OBP2A/OBP2B are both Lipocalin-family proteins. Specific recurrent mutations are labeled above the dots (OBP2A: R82Q/W; OBP2B: G126E/*/R). The highlighted spheres in cartoon structure along the protein chains visualized in (A) indicated mutated residues that changed properties screening out from (B), as shown by the two color coded residues onto the surface and binding pocket. Color code (A): The main chain of OBP2A is in grey; The main chain of OBP2B is in light brown. The hydrophilic residues are in red; The hydrophobic residues are in magenta. Color code (B): splice mutations are in orange dots; missense mutations are in green dots; in-frame deletions are in brown dots.

**Future Perspectives and applications of OBPs**

While the exact biological function of OBPs in human olfaction remains largely unknown, OBPs and lipocalins in general remain interesting candidates for a variety of biomimetic applications. Their size and robustness provide ideal conditions to explore their usage in molecular detection devices, such as artificial noses and biomarker sensors, or as drug and flavor delivery agents. Current efforts in OBP research are mostly directed on applications related to their putative function in the olfactory sense (Zahra et al., 2024), thereby focusing on increasing their specificity for targeted air-quality monitoring, food quality assessment, and medical diagnostics (Hellmann, 2020). Over the past decades, several studies emerged with the aim to fabricate individual sensors, tuned to identify and detect volatile organic compounds with high specificity (Pelosi et al., 2018a; Hurot et al., 2019, 2020). However, until now most electronic noses use only poor specificity gas sensors based on metal oxides and conducting polymers (Zhai et al., 2024). The major drawback of these gas-sensing materials is doubtless their low selectivity and stability. Recent trends in gas-phase biosensing showed an opening towards different protein-based elements: odorants receptors, OBPs, and odorant receptor peptides (Barbosa et al., 2018). Hereby, OBPs stand out as suitable detecting elements for artificial gas and smell sensing (Pelosi et al., 2018b). In contrary to membrane-bound receptors which, despite their high-selectivity (Li et al., 2016; Haag and Krautwurst, 2022a), are highly unstable and sensitive to external stress, and odorant receptor peptides for which only a very limited number of suitable peptides is known for chemical sensing, OBPs are extremely stable to high temperature, refractory to proteolysis and resistant to organic solvents (Pelosi et al., 2018b; Haag and Krautwurst, 2022b). Preliminary studies showed that porcine OBP can be tailored towards higher specificity to distinguish between carvone enantiomers (Paesani et al., 2025). Fine-tuning the binding properties of OBPs and lipocalins towards the detection of specific molecules or molecular families provides a stepping stone for artificial sensor devices and can open up applications in the biomedical field (Mulla et al., 2015; Pelosi et al., 2018a; Haag and Krautwurst, 2022a), leveraging OBPs to detect biomarkers in the breath of patients to diagnose diseases (e.g. hexanal in lung cancer (Mousazadeh et al., 2022)). Additionally, machine learning algorithms are being employed to analyze the complex data generated by these OBP-based sensors, enabling improved odor recognition patterns and enhancing the overall performance of artificial olfactory systems (Lötsch et al., 2019; Hellmann, 2020).

Closely linked to olfactory research is flavor research; which combines taste (gustation) and smell (olfaction). In flavor research, further exploring the transporter functions of OBPs to improve the properties of flavor distribution in different food matrices could be an interesting opportunity (Boichot et al., 2022). OBPs may be leveraged to encapsulate small odor and flavor molecules, thus preventing them from reaching taste receptors too fast (Ren et al., 2024). They could be fine-tuned towards "masking" odors or towards reducing the effect and perception of specific odor and flavor molecules (Gascon, 2007). Similarly, OBPs can be targeted to transport drugs to specific locations inside the body (Choi et al., 2022) or designed to target specific cells or tissues (Nästle et al., 2023). A benefit of leveraging these proteins as drug carriers is their limited immune response as direct members of the organism (Yang et al., 2023). Their potential to seize the harmful compounds make them favorable candidates for applications in detoxification (Grolli et al., 2006; Nakanishi et al., 2024). Thus, due to their ability to bind and transport hydrophobic molecules, OBPs and lipocalins may be engineered towards different applications that go beyond their putative roles in the olfactory sense. Finally, recent studies on the aggregation of bOBPs, highlighted the usefulness of leveraging the properties of hydrophobic beta-barrel to study fibrillogenesis in neurodegenerative diseases (Stepanenko et al., 2023, Sulatskaya et al., 2024). This could provide another exciting domain for OBPs in bio-chemical applications and lead to the development of new therapeutic strategies to prevent pathological beta-barrel aggregation.

## Conclusion

In this systematic review, we showed that despite their fame in context with the olfactory sense, OBPs are overexpressed in various tissues throughout the human body, especially in the male and female reproductive organs. Based on this observation and the ability of OBPs to bind hydrophobic compounds, we hypothesized their potential function as hormone transporters. Based on their overexpression in the reproductive organs, we further explored their potential implications in cancer tissues of the reproductive organs, since several cancer types are hormone dependent. While our review clearly showed that human OBPs are overexpressed in both healthy tissues, the statistical analysis using CIBRA to evaluate the impact of the underlying genomic alterations in cancer tissues remained inconclusive. Still, it cannot be ruled out with certainty that OBP could function as hormone transporter and participate in cancer progression and tumor genesis. To verify this, more advanced experiments are needed, for instance using expression profiling and knockouts in cancer cells, possibly with a focus on ovarian and breast cancer. Overall, integrating the study of OBPs with other members of the lipocalin family and possibly other olfactory proteins such as ORs, could help our understanding of the diverse functions of OBPs in the human body. In future studies, computational simulations from molecular dynamics may provide additional insight on the structure and dynamics of hOBPs and help to characterize their uptake mechanisms and affinities towards odorant and flavor molecules. Hereby, the availability of available crystal structures of OBPs in their apo and holo form with different ligands (Pevsner et al., 1990; Ramoni et al., 2001; Schiefner et al., 2015), will be crucial for rational simulation studies to improve their selectivity and physical properties. On the experimental side, techniques such as fluorescence binding assays (D’Onofrio et al., 2020), circular dichroism spectroscopy (Stepanenko et al., 2024b), and isothermal titration calorimetry have been widely employed to characterize protein–ligand interactions (Moitrier et al., 2022). These methods provide crucial quantitative and structural insights that complement simulation-based predictions, enabling validation of binding affinities and conformational changes upon active site binding.

## METHODS

### 1. Structured Literature Overview

To provide a structured overview of OBPs, their function, characteristics and potential applications, we systematically searched PubMed, Scopus, and Web of Science databases for publications related to OBP2A/2B gene expression and their relevance in reproductive cancers. The search covered all available literature without any restriction on the publishing date. The provided references cover publications from 1982-2025, using combinations of the terms: "OBP2A/2B", "olfactory binding proteins", "ovarian cancer", "breast cancer" "uterine cancer" and "gene expression", connected by Boolean operators (AND, OR). Only English-language, peer-reviewed original research articles were included. While the focus of the work is on human olfaction and tissues, we mention non-human examples and studies from insects and mammals when needed to get a full overview of the problem. The authors screened titles and abstracts independently, followed by full-text screening and discussed the results and main findings together.

### 2. Metadata Analysis

Figure 10 shows a schematic overview of the meta-analysis approach using research questions from the structured literature overview.

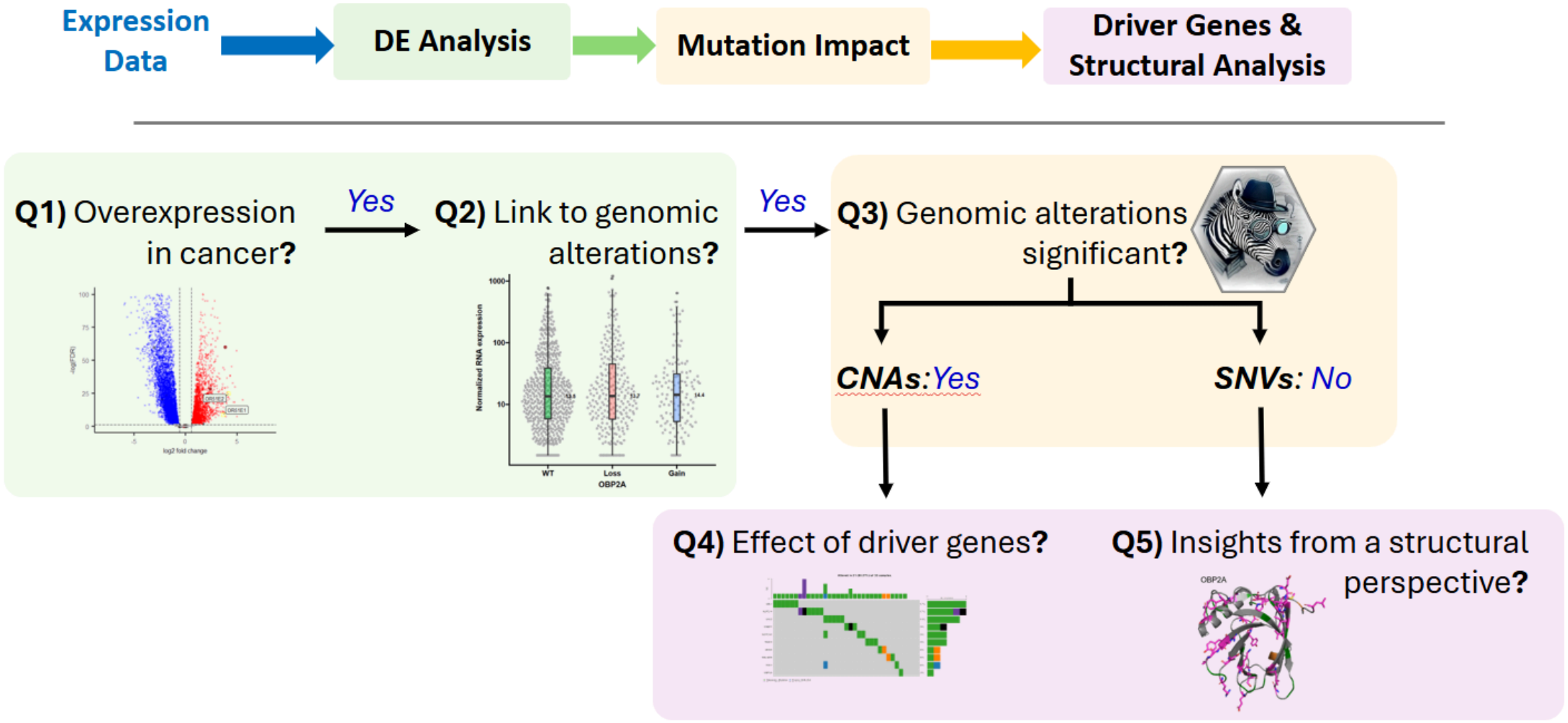

**Figure 10. Schematic overview of the metadata analysis approach to systematically assess additional putative functions of OBPs outside of olfaction.** The RNA-seq data of *OBP2A* and *OBP2B* genes are used as inputs for differential expression analysis (DE Analysis) to verify the overexpression in cancer tissues of reproductive organs (Q1), connecting expression patterns with genomic alterations (Q2), look into mutation impacts of CNAs and SNVs (Q3), and lastly, since the genomic alterations CNAs and SNVs turned out to be significant and not significant, respectively, we conclude our analysis with looking into the effects of driver genes and an in-depth structural analysis of OBP2A/OBP2B and their mutational variants (Q4 and Q5).

Starting from RNA-seq expression data to (Q1) determine whether OBPs are overexpressed in any cancer types, (Q2) to confirm links to underlying genomic alteration, (Q3) verify their significance, (Q4) check if the overexpression of OBP2A and OBP2B genes are caused by neighboring driver genes of reproductive cancers that are located on the same locus (chromosome 9) as OBPs, and (Q5) gain insight from a structural perspective by leveraging MSA and impact of identified structural mutants. The technical details for each of the individual steps is provided below. We first performed differential gene expression analysis of OBPs in several cancer types, with the focus on reproductive tissues: ovarian, breast, prostate and uterine cancer. To further understand the underlying mutational profile in terms of cancer related copy number alterations (CNAs) and single nucleotide variants (SNVs), we used CIBRA, a computational method to assess the effect of these genomic alterations on system-wide gene expression (Lakbir et al., 2024) along with MSA and structural variants analysis for full mutational overview.

### 2.1 Databases

#### 2.2.1 HPA, PaxDB and PDC

The data is retrieved from the Human Protein Atlas (HPA), Protein Abundance Exchange Database (PaxDB) and Proteomic Data Commons (PDC). Queried series generated based on clinical data of OBP2A and OBP2B in various cancers including ovarian, breast, uterine, prostate, melanoma, lung and colorectal cancers.

#### 2.2.2 The Cancer Genome Atlas (TCGA) data

Clinical information and mutation calls for TCGA were obtained from cBioportal. Gene expression data quantified as counts were retrieved from the ovarian, breast, prostate, uterine, melanoma, lung, colorectal cancers (harmonized with GTEx), The RNA-Seq data has been normalized by calculating the Transcripts Per Kilobase Million (TPM) or Fragments Per Kilobase Million (FPKM).

## 3. Analysis of sequences, expression, mutation significance, and driver genes

### 3.1 Multiple Sequence Alignment (MSA)

The MSA was performed with Clustal Omega(Sievers et al., 2011) with default parameters using the Job Dispatcher from the European Bioinformatics Institute (EMBL-EBI). Protein sequences from human lipocalins and mammalian OBPs were retrieved from UniProt. The MSA was conducted with these protein sequences and visualized using JalView (Troshin et al., 2011, 2018) with default parameters. The correlation identity was generated using the NCBI BLAST (Basic Local Alignment Search Tool)(Altschul et al., 1990; Camacho et al., 2023) with corresponding protein sequences from uniprot.

### 3.2 Differential Expression Analysis (DEA)

DE analysis was performed using the R package limma (version 3.28.14, default parameters) and edgeR (3.14.0 version, default parameters) following the protocol from Chen *et al* (Chen and MacDonald, 2022). Harmonized and normalized gene counts were retrieved for TCGA and GTEx from the UCSC Toil Recompute Compendium (Lonsdale et al., 2013). Genes with counts $< 10$ or no variance were filtered from the analysis. The results were subsequently visualized using volcano plots using EnhancedVolcano R package (version 1.27.0) (Blighe, 2018). To estimate expected lipocalin exon mutations across cancer types, we calculated a background mutation rate using the most mutated cancer type, applied it to lipocalin exon lengths, and scaled the result by the number of samples per cancer type.

### 3.3 Computational Identification of Biologically Relevant Alterations (CIBRA)

CIBRA is accurate for full overview of alterations that even have low frequency but engage high potential related to cancers. It is an accurate way of detecting somatic alterations based on genome instability caused for cancer. To identify the gene alterations, CIBRA shows in significant scores that system-wide impact. Genes with an adjusted p-value (FDR) of $< 0.05$ were considered significantly differentially expressed. The log2 fold change (log2FC) was used to determine the direction and magnitude of expression changes. This enables systematically involved biological relevance of targeted genes with certain tumorigenesis. Boxplots were generated to compare the normalized expression levels of OBP2A and OBP2B among WT, gain and loss. These levels are based on copy-number analysis, and indicates as resigned matrix that -2 as deep deletion, is recognized to be a homozygous deletion; -1 as shallow deletion indicates a loss, is possibly a heterozygous; o is diploid. 1 as gain and indicates a few additional copies, which often broad wide. 2 as amplification shows more copies.

### 4. Data availability

The entire analysis workflow, including dataset processing scripts and visualizations, has been made accessible via GitHub at https://github.com/CMFTIME7/OBP-cancer.

**Acknowledgements**

This publication is part of the project ODORWISE with file number OCENW.M.23.020 of the research programme Open Competition Domain Science M23-1 which is (partly) financed by the Dutch Research Council (NWO). The authors would like to thank Ximeng Wang and Liu Yang for their contributions within their literature overview and M.Sc. research project, respectively. H.M. and M.C. would like to thank the Chinese Scholarship Council for funds (personal PhD grant M.C.) and the Giract Foundation for a 1st year PhD fellowship (personal grant M.C.). The authors acknowledge the SURFsara compute cluster hosted by SURF and the BAZIS research cluster hosted by VU for the computational time and the provided technical support.

**Author contributions statement**

All authors were involved in the work, M.C. and S.L. contributed equally to the work. M.C., S.L. performed the data analyses and interpretation; M.C. prepared the original draft of the manuscript and completed the literature search and literature overview; S.L. provided supervision for B.Sc. and M.Sc. students (V.M. and M.J.); M.J. and V.M. performed the DGE analyses, the initial literature search and overview; S.A. and H.M. conceptualized the project idea, H.M. was responsible for funding and main supervision; S.A. and S.L. provided critical feedback. All authors reviewed and approved the final version of the manuscript.

**Additional information**

**Supporting Information.** The online version contains supplementary material available online, the code for the analysis described in the methods section is available on GitHub (see Section 4 "Data Availability").

**Competing interests.** The authors declare no competing interests.

# Electronic Supporting Information

## Beyond Olfaction: New Insights into Human Odorant Binding Proteins

Mifen Chen[a], Soufyan Lakbir[a][b], Mihyeon Jeon[a], Vojta Mazur[a], Sanne Abeln[b], and Halima Mouhib*[a]

[a] Department of Computer Science, VU Bioinformatics Group, Vrije Universiteit Amsterdam, De Boelelaan 1105, 1081 HV Amsterdam, The Netherlands.
[b] Department of Computer Science, AI Technology for Life, Universiteit Utrecht, Heidelberglaan 8, 3584 CS Utrecht, The Netherlands

*corresponding author: h.mouhib@vu.nl

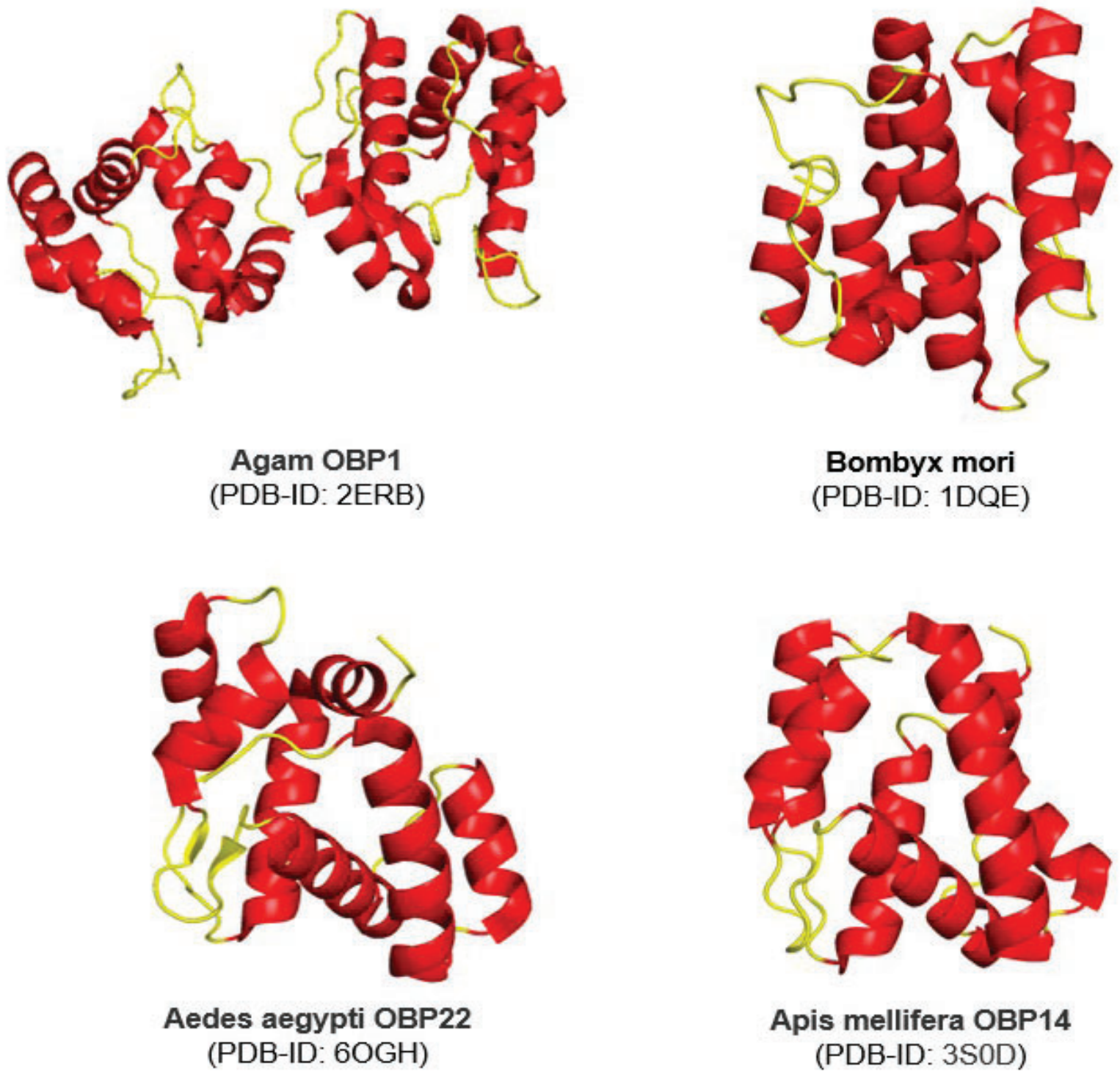

**Figure S1.** Insect OBPs structures: helices (red), coils (yellow). It should be noted that Agam OBP1 (PDB-ID: 2ERB) is reported as a dimer, indicating molecules passing through channels to both chains, while all others are reported as monomers.

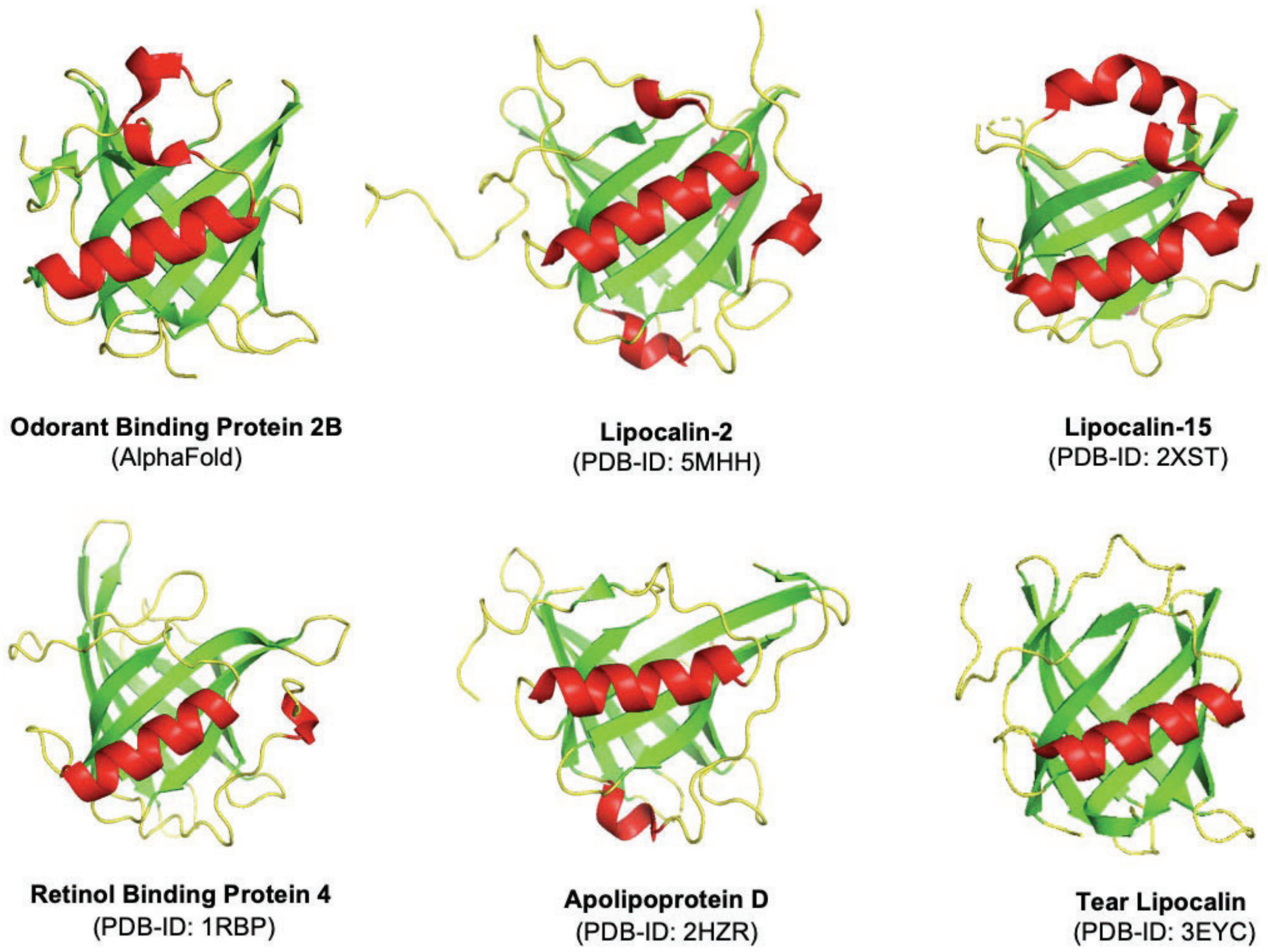

**Figure S2.** Protein structures of different human lipocalins. The colors red, yellow, green depict the alpha helices, loops, and beta-sheets of the structures, respectively. Note that the structure of Odorant Binding Protein 2B was generated using AlphaFold. For the experimental structure of OBP2A see Figure 1 in the manuscript.

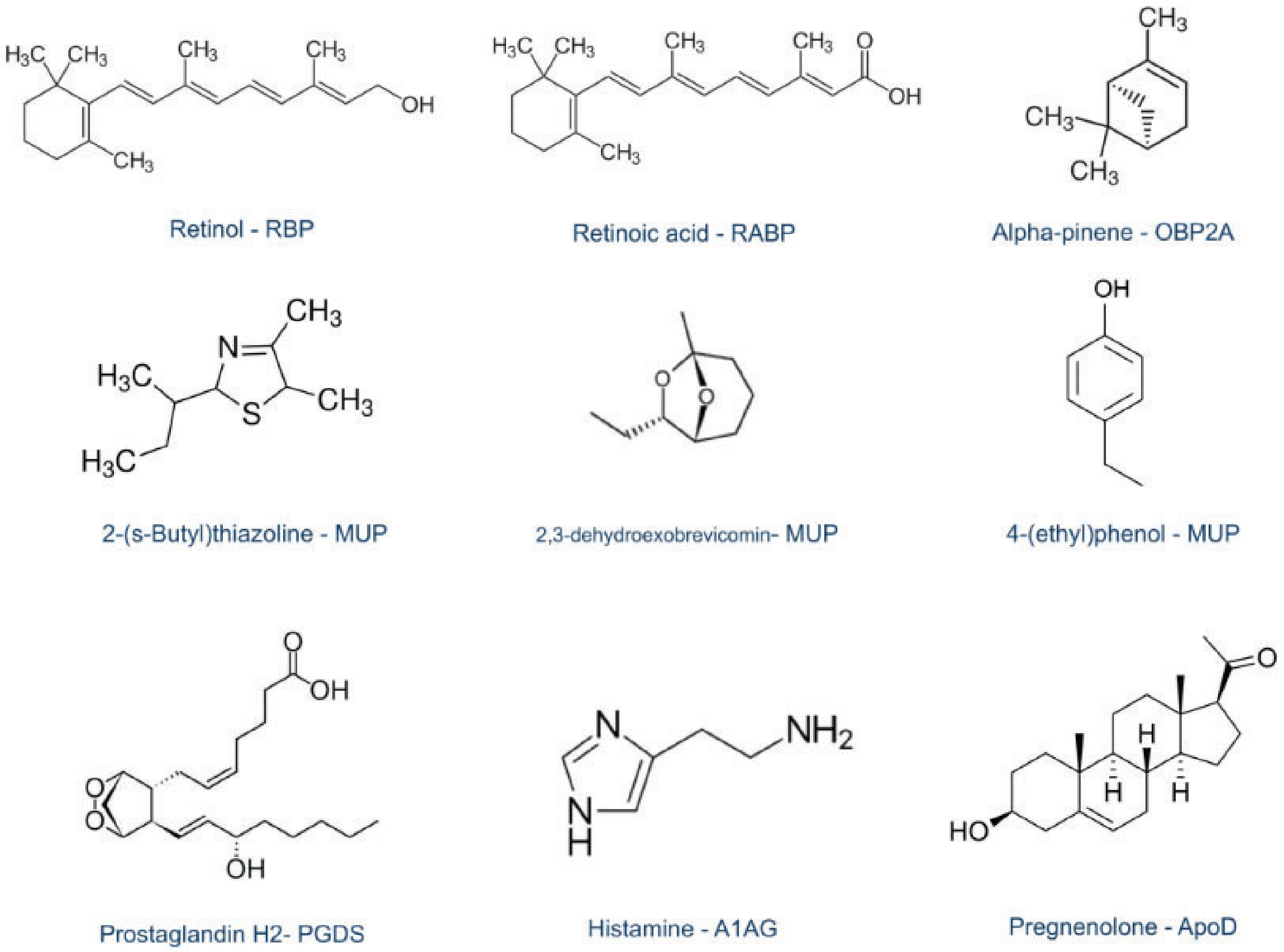

**Figure S3.** Chemical formulas of known ligands bound by lipocalins. Beneath each formula, the corresponding lipocalin protein and the name of the ligand are listed.

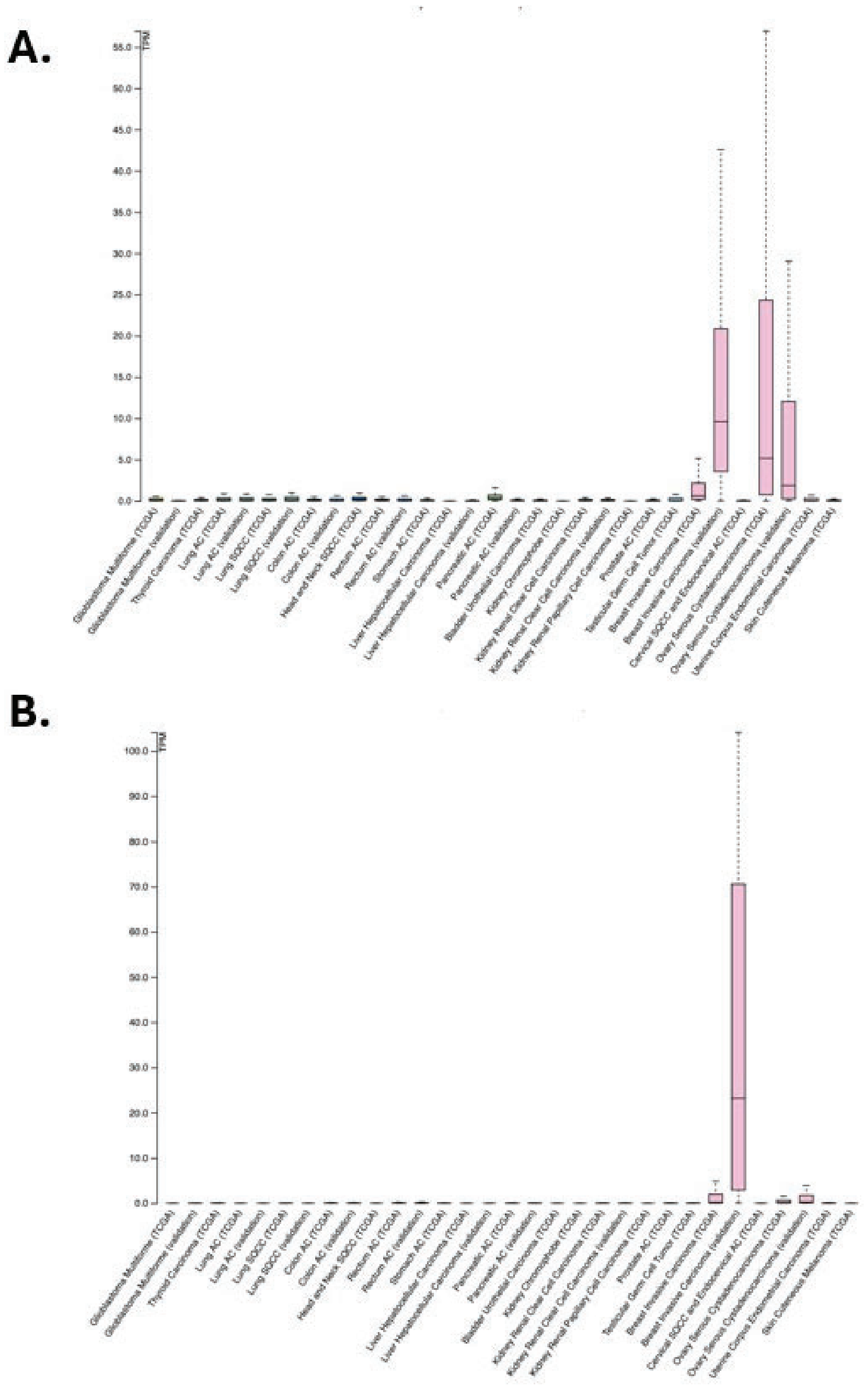

**Figure S4-1. RNA and protein expression distribution of OBP2A and OBP2B across various cancer tissues from TCGA dataset.** RNA expression of OBP2A (A) and OBP2B (B). The X axis presents cancer types. The Y axis displays Transcripts Per Kilobase Million (TPM). The central line within each box represents the median. The outlier data points that differ significantly from the rest are shown in the dots. Pink-colored boxes indicate high expression levels.

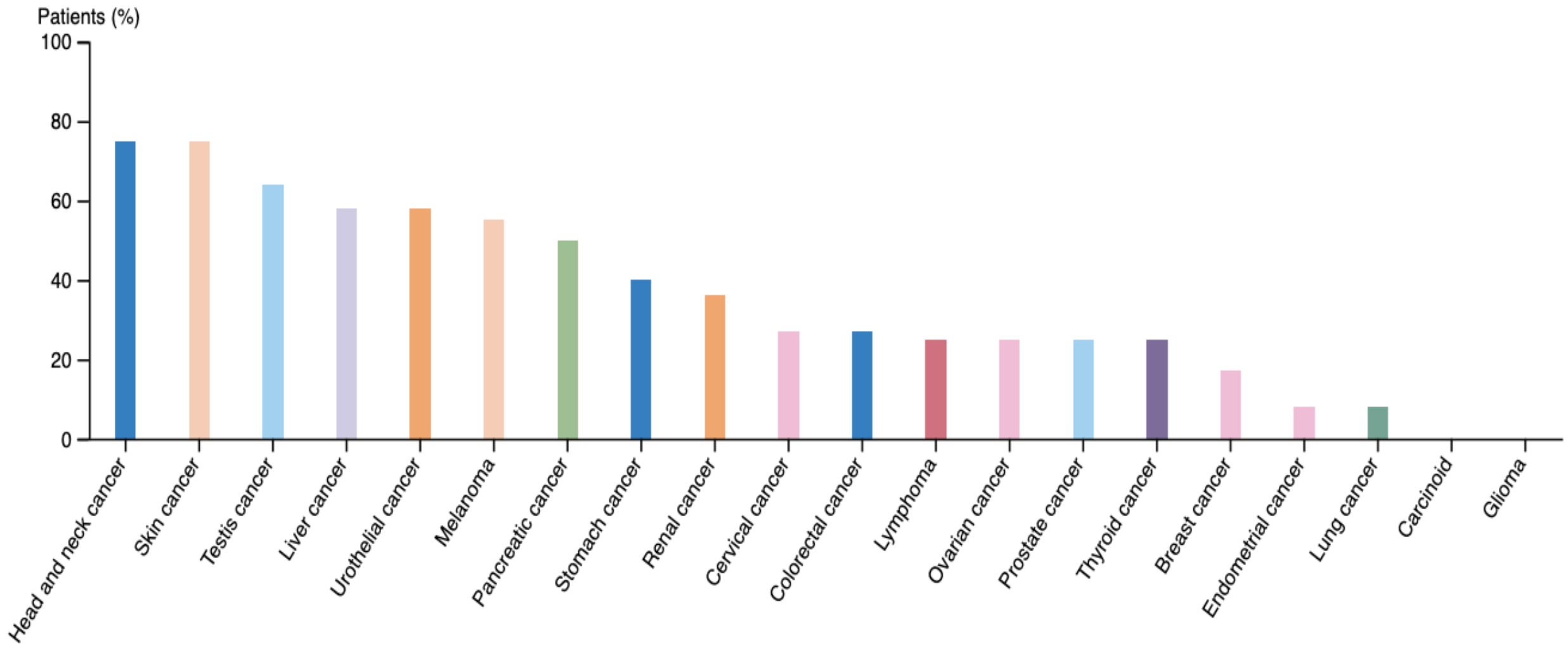

**Figure S4-2. RNA and protein expression distribution of OBP2A and OBP2B across various cancer tissues from TCGA dataset.** Protein expression of OBP2A (not available for OBP2B on the HPA). The X axis shows several cancer types. The Y axis presents the percentage of patients. Each box is color-coded corresponding to each cancer type.

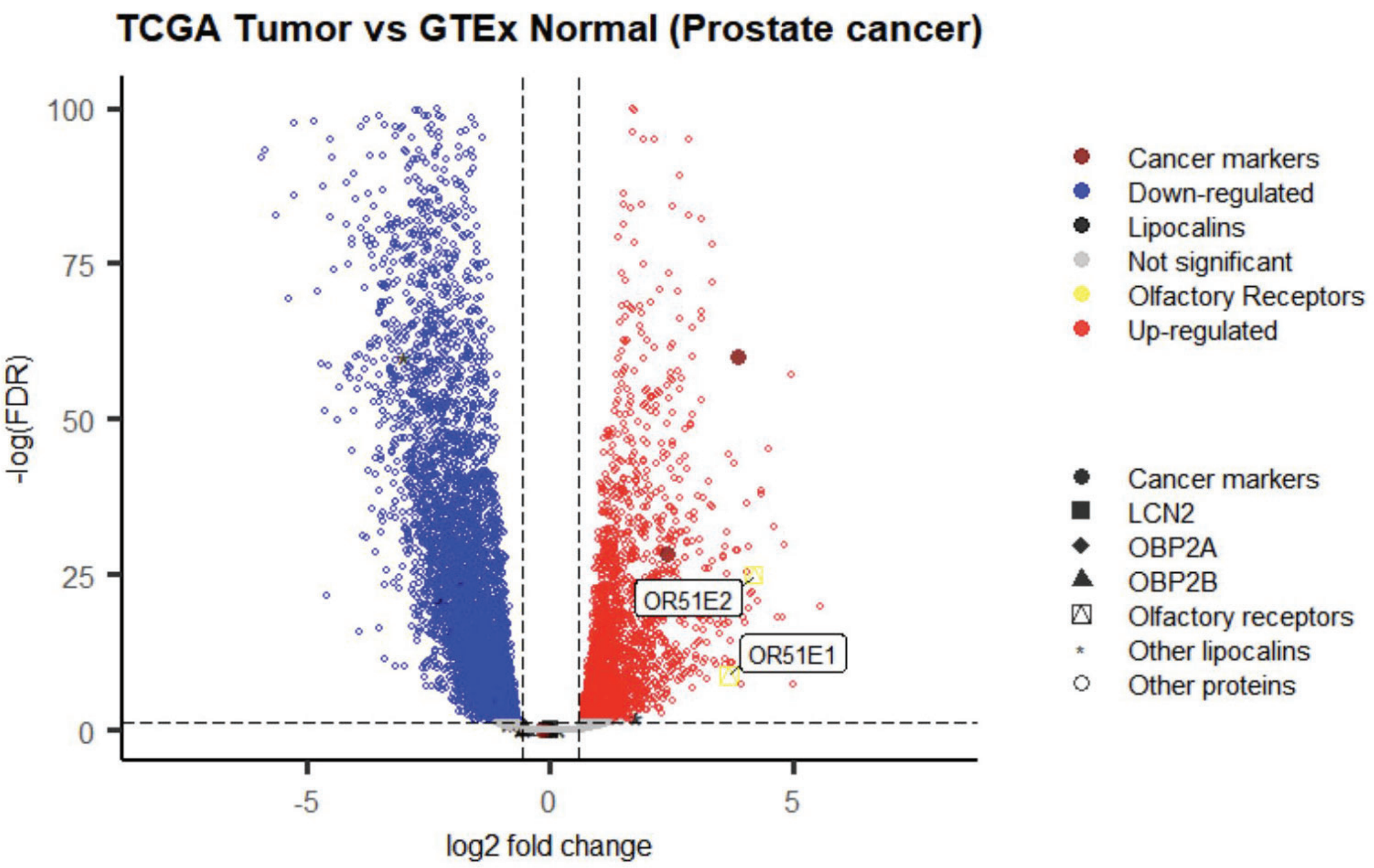

**Figure S5: Differential Gene Expression Analysis of Odorant Receptors in Prostate Cancer. Volcano plots display the differential gene expression between normal and tumor samples in prostate cancer.** The x-axis represents the log2 fold change in gene expression between cancer and healthy tissues, while the y-axis represents -log(False discovery rate), indicating statistical significance. The same human lipocalins and cancer markers are labeled as in Figure 6 of the manuscript, with genes colored blue for down-regulated and red for up-regulated. OR51E2 and OR51E1 genes were upregulated while OR2T6 and OR5B21 were filtered out due to their low counts.

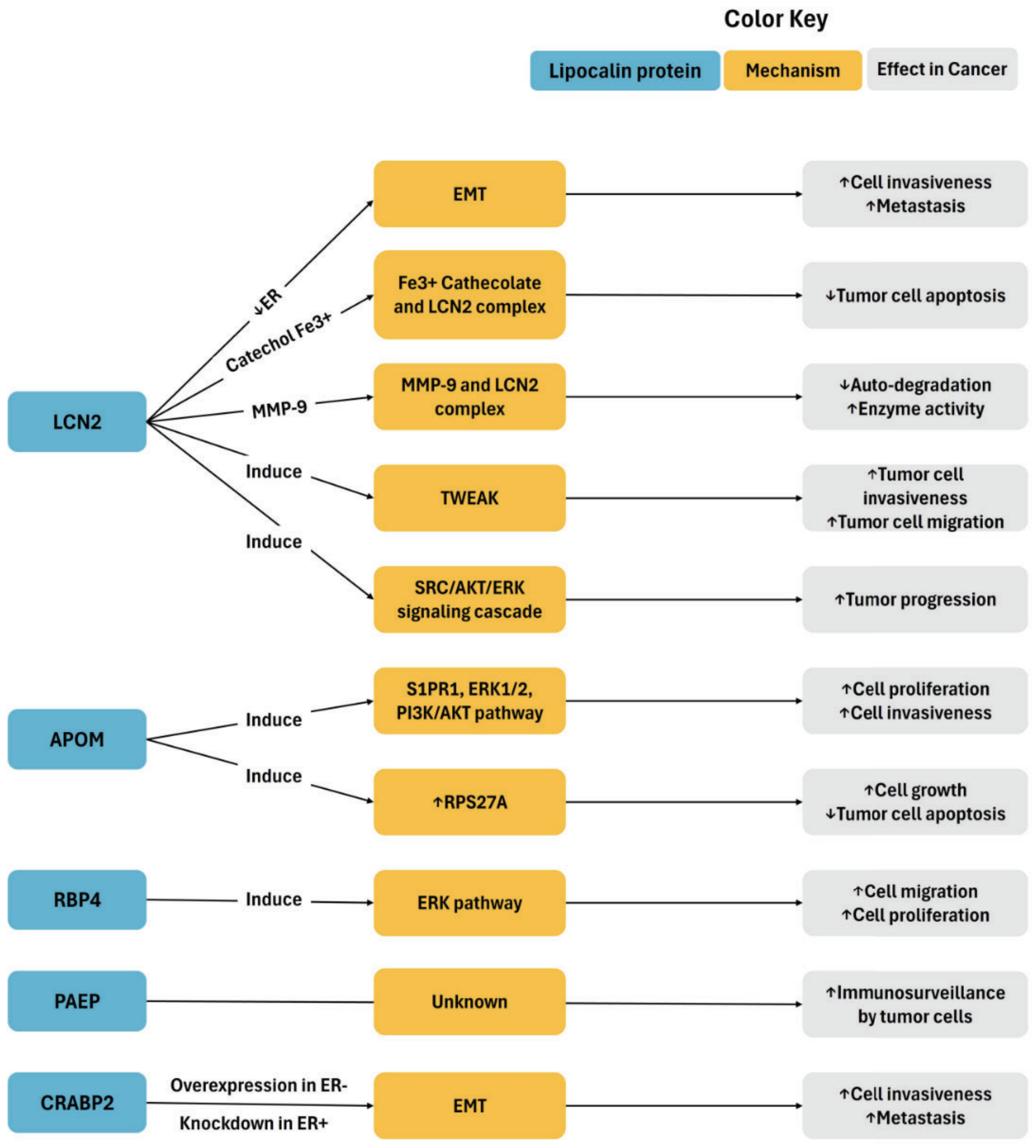

**Figure S6. Overview of the known mechanisms and effects of lipocalins in cancer.** The figure provides an overview of selected members (LCN2, APOM, RBP4, PAEP and CRABP2) of lipocalin family proteins that contribute to cancer progression. It highlights the key pathways and reactions of them in the tumorigenesis, including up and down regulation on each track. Remarkably, the mechanism by which PAEP influences cancer remains to be elucidated.

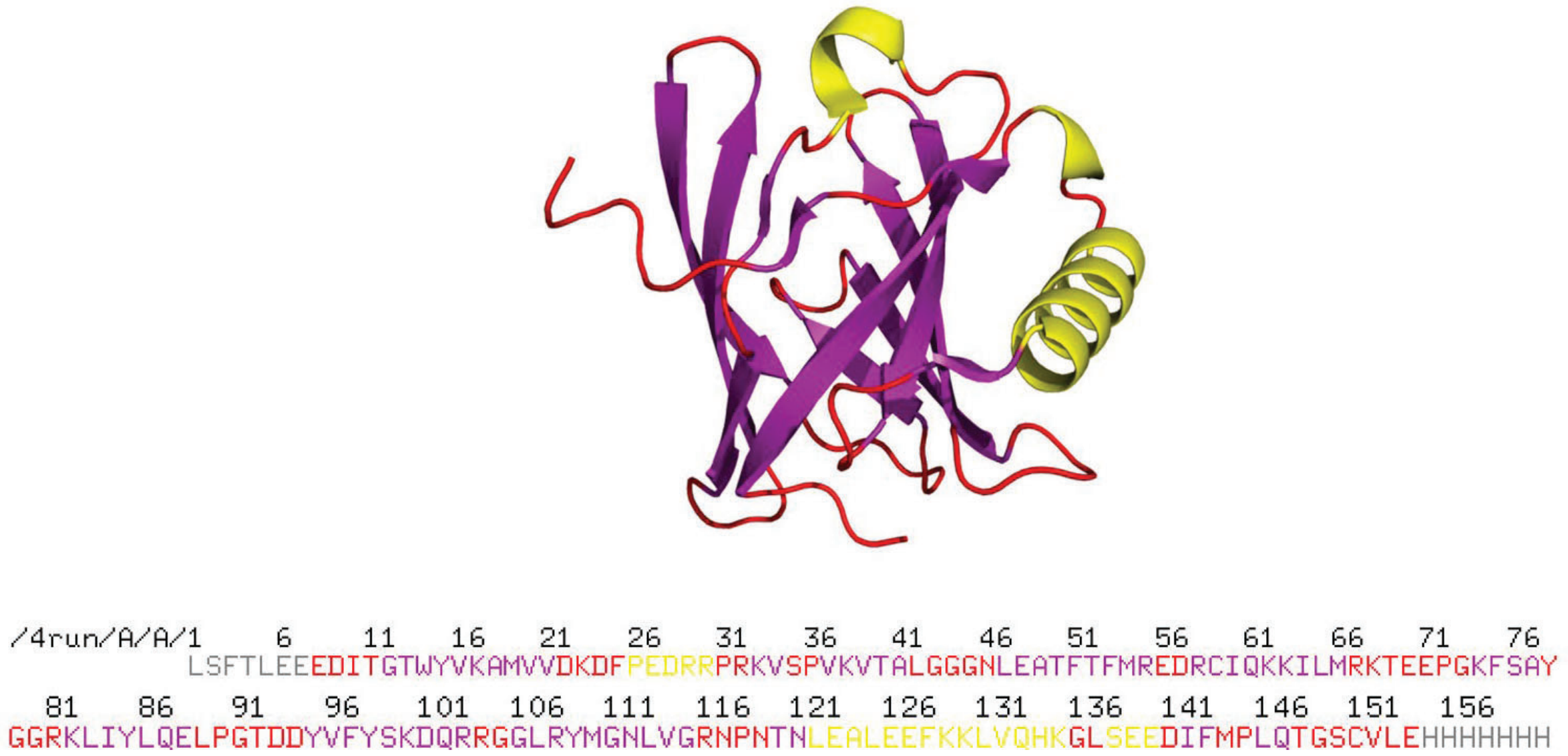

**Figure S7. Beta barrel and alpha helix displayed in sequences on crystal structure of OBP2A (PDB-ID: 4RUN) to show in regions.** Color-coded: red (coils and loops); Purple (beta-barrel); yellow (alpha-helix).

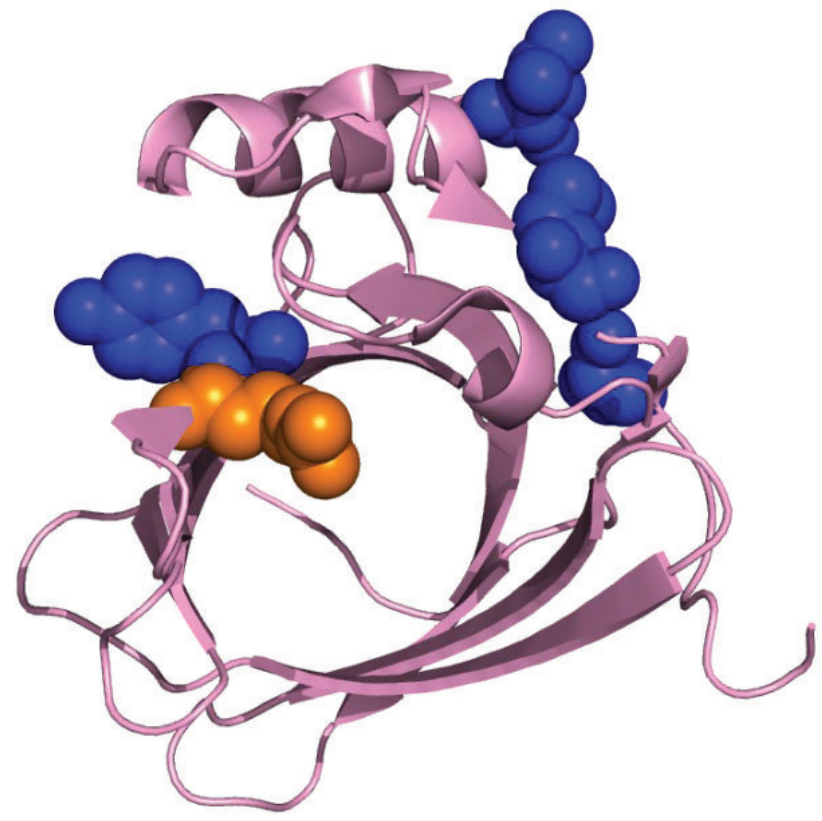

**Figure S8. Visualization of different residues between OBP2A and OBP2B on the structure of OBP2A (PDB-ID: 4RUN).** The five highlighted residues between OBP2A and OBP2B shift their properties from hydrophobic to hydrophilic or vice versa. Hydrophilic residues (R108) are highlighted in orange, hydrophobic residues (Y109, P118, L122, M144) are highlighted in blue (see also Table S2 for full information).

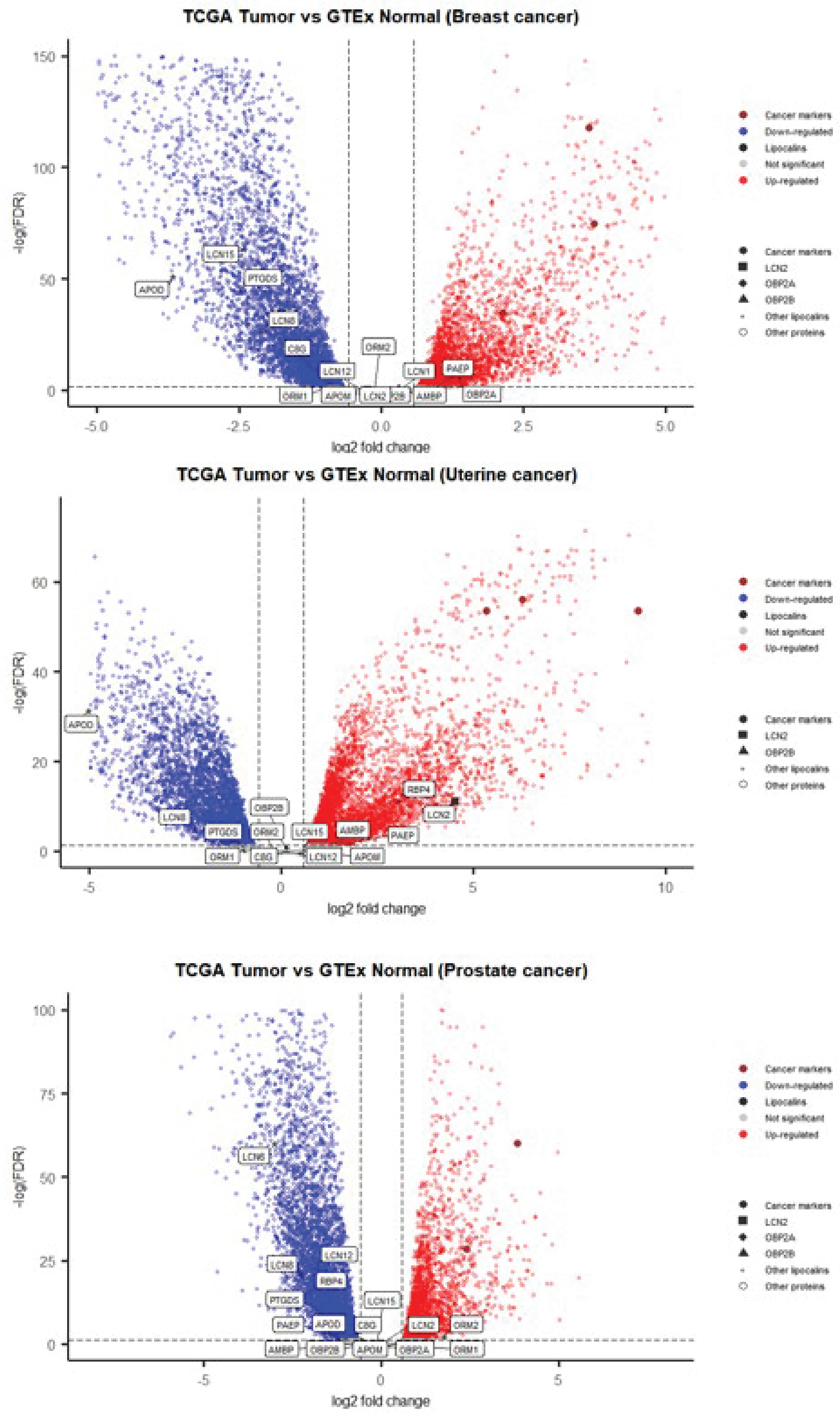

**Figure S9-1.** Differential Gene Expression Analysis of OBP2A, OBP2B, human lipocalins in breast, uterine, and prostate (cf. Figure S9-2 for full description).

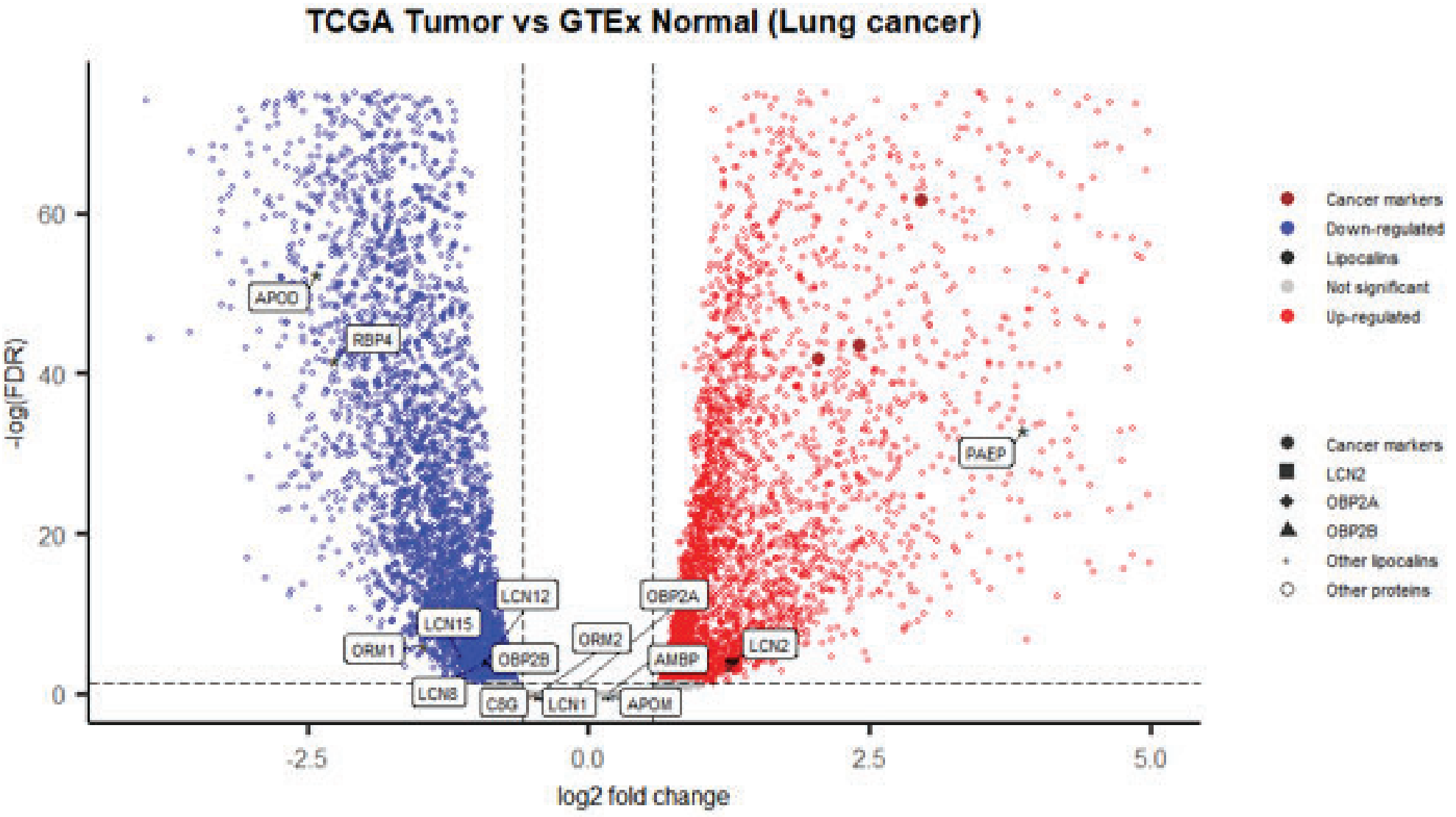

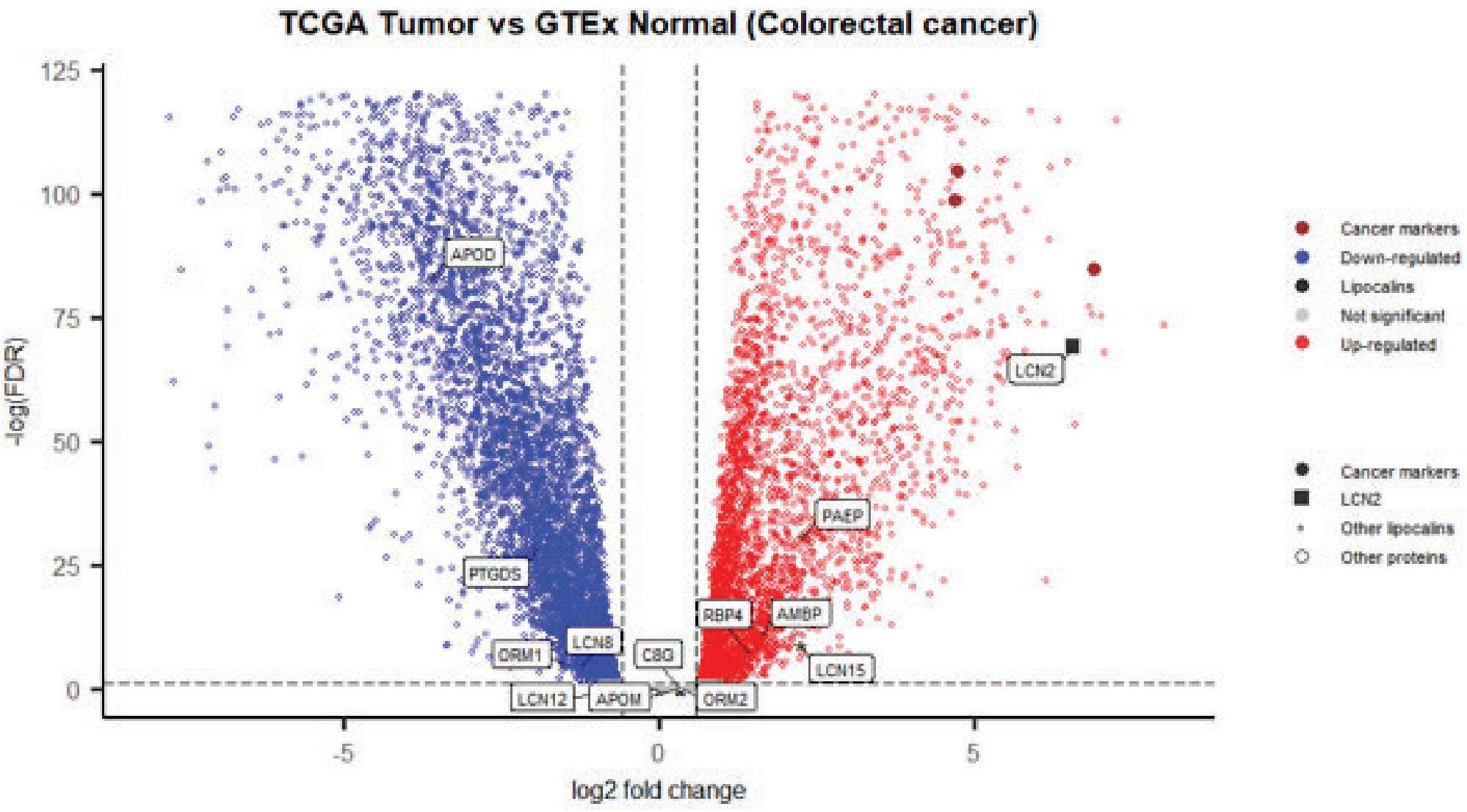

**Figure S9-2.** Differential Gene Expression Analysis of OBP2A, OBP2B, human lipocalins in lung and colorectal cancer. Volcano plots display the differential gene expression between normal and tumor samples in five cancer types. The x-axis represents the log2 fold change in gene expression between cancer and healthy tissues, while the y-axis represents -log(False discovery rate), indicating statistical significance. The same human lipocalins and cancer markers are labeled as in Figure 6 of the manuscript, with genes colored blue for down-regulated and red for up-regulated.

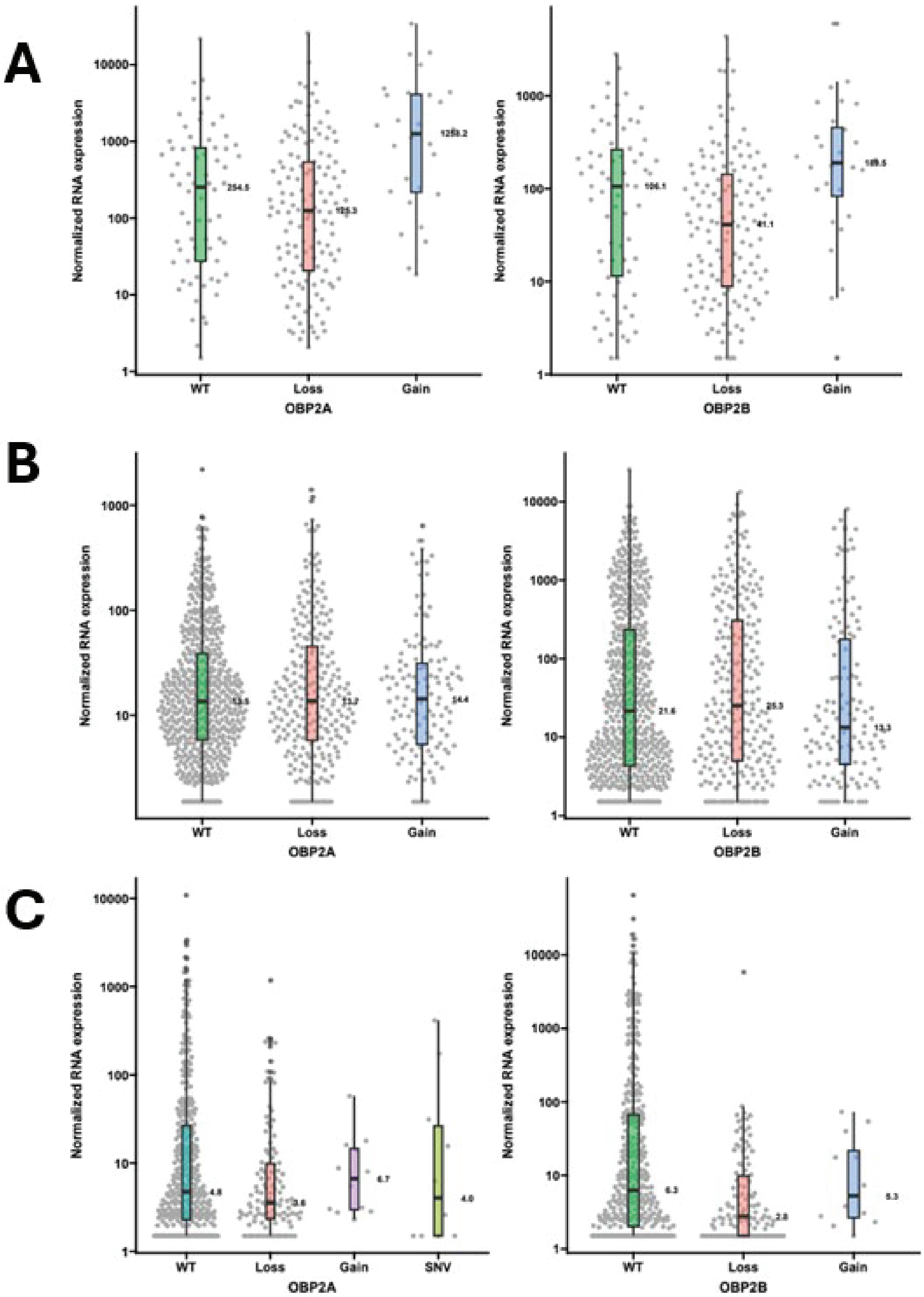

**Figure S10-1.** Normalized RNA expression of OBP2A and OBP2B genes in (A) ovarian cancer, (B) breast cancer, (C) uterine cancer. See continued caption of Figure S10-2 for full description.

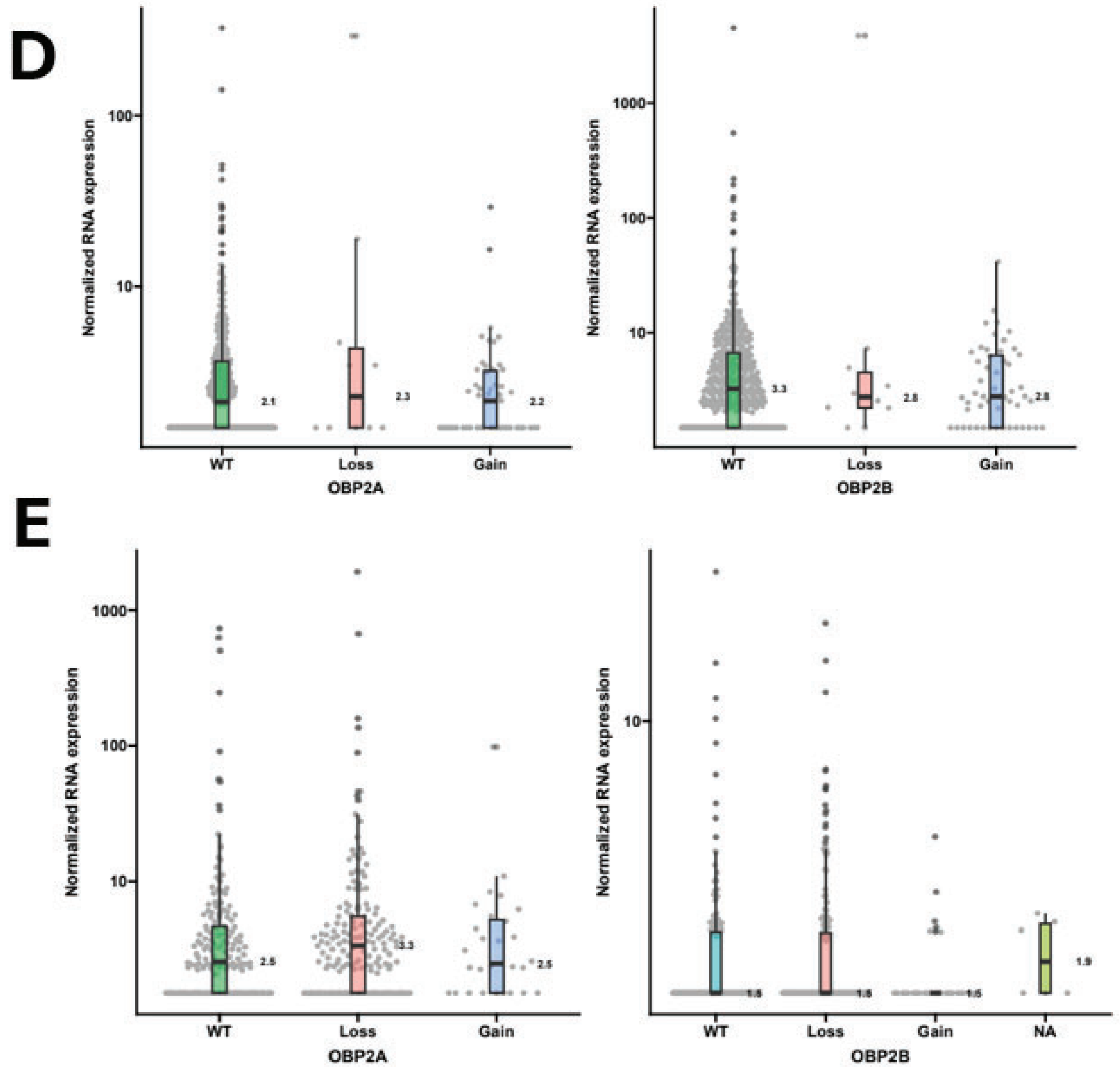

**Figure S10-2. Normalized RNA expression of OBP2A and OBP2B genes in (D) prostate cancer, and (E) melanoma cancer.** The boxplots compare normalized RNA expression levels across different cancer types in different categories for the genes *OBP2A* (left) and *OBP2B* (right). In X-axis, the categories include: "WT" (wild-type), "Loss" (copy number loss), and "Gain" (copy number gain). WT (Wild-Type): Represents tumors without mutations in OBP2A or OBP2B. Loss: Represents tumors where OBP2A or OBP2B has a deletion (loss of function). Gain: Represents tumors where OBP2A or OBP2B is amplified (increased expression). SNV (in uterine cancer): Represents tumors with single-nucleotide variants (point mutations) in OBP2A. Colors indication: Green (WT); Red (Loss); Blue (Gain). Y-axis is Normalized RNA expression (on a logarithmic scale). The spread of data points (gray dots) within the boxplots indicates the variability in normalized expression level across samples in each category. The wider spread suggests more variability in RNA expression among these samples, which has narrower variability. There are several outlier observations visible as points beyond the boxplot whiskers, especially in the WT category, highlighting some samples with exceptionally high expression levels. Some points in each category fall outside the whiskers, indicating potential extreme values in the dataset. Displayed median values in bold inside the boxes.

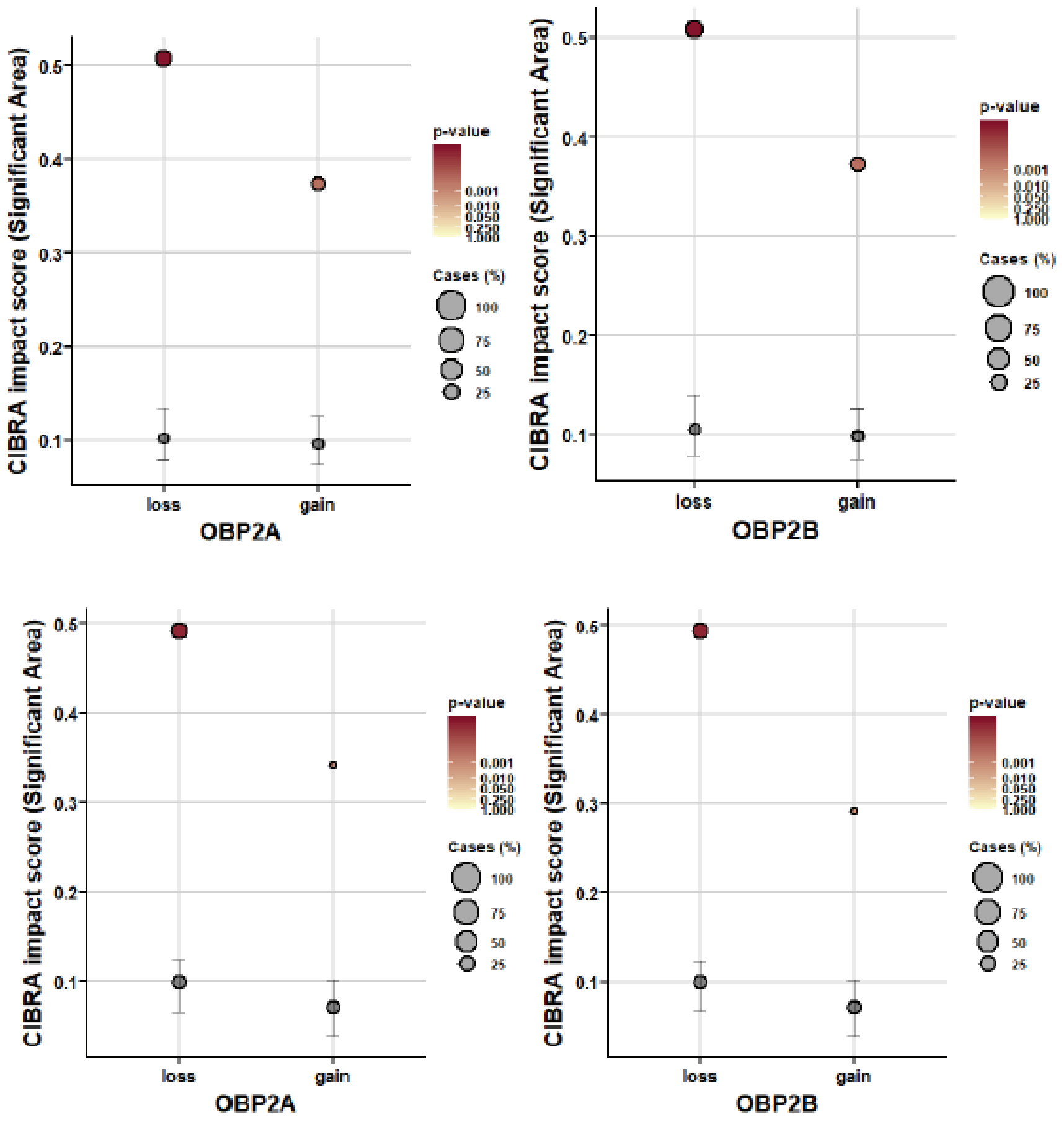

**Figure S11.** **CIBRA impact scores for OBP2A and OBP2B copy number alterations (CNA) across breast (upper), and uterine cancers (below).** This figure illustrates the CIBRA impact scores for loss and gain events in *OBP2A* and *OBP2B* across breast cancer (upper part of the figure) and uterine cancer (lower part of the figure). Each bubble represents the significant area of impact scores associated with either copy number loss or gain, as analyzed by the CIBRA pipeline. The size of the bubbles indicates the percentage of cases (%) affected by each alteration, and the color gradient corresponds to the p-value, where darker red shades denote stronger statistical significance (lower p-values); Light red suggest less significant results (higher p-values). The vertical axis represents the CIBRA impact score (significant area), which reflects the extent to which each CNA (loss or gain) influences the molecular profile in these cancers; higher values suggest a greater impact of the genetic alteration. For X-axis, loss: tumors where the gene is deleted (reduced copy number); gain: tumors where the gene is amplified (increase copy number). Error bars represent the interquartile range (IQR) of the data. A loss implies shallow deletion which could be a heterozygous deletion. The gain indicates a low level of additional copies, which is usually in a wide range.

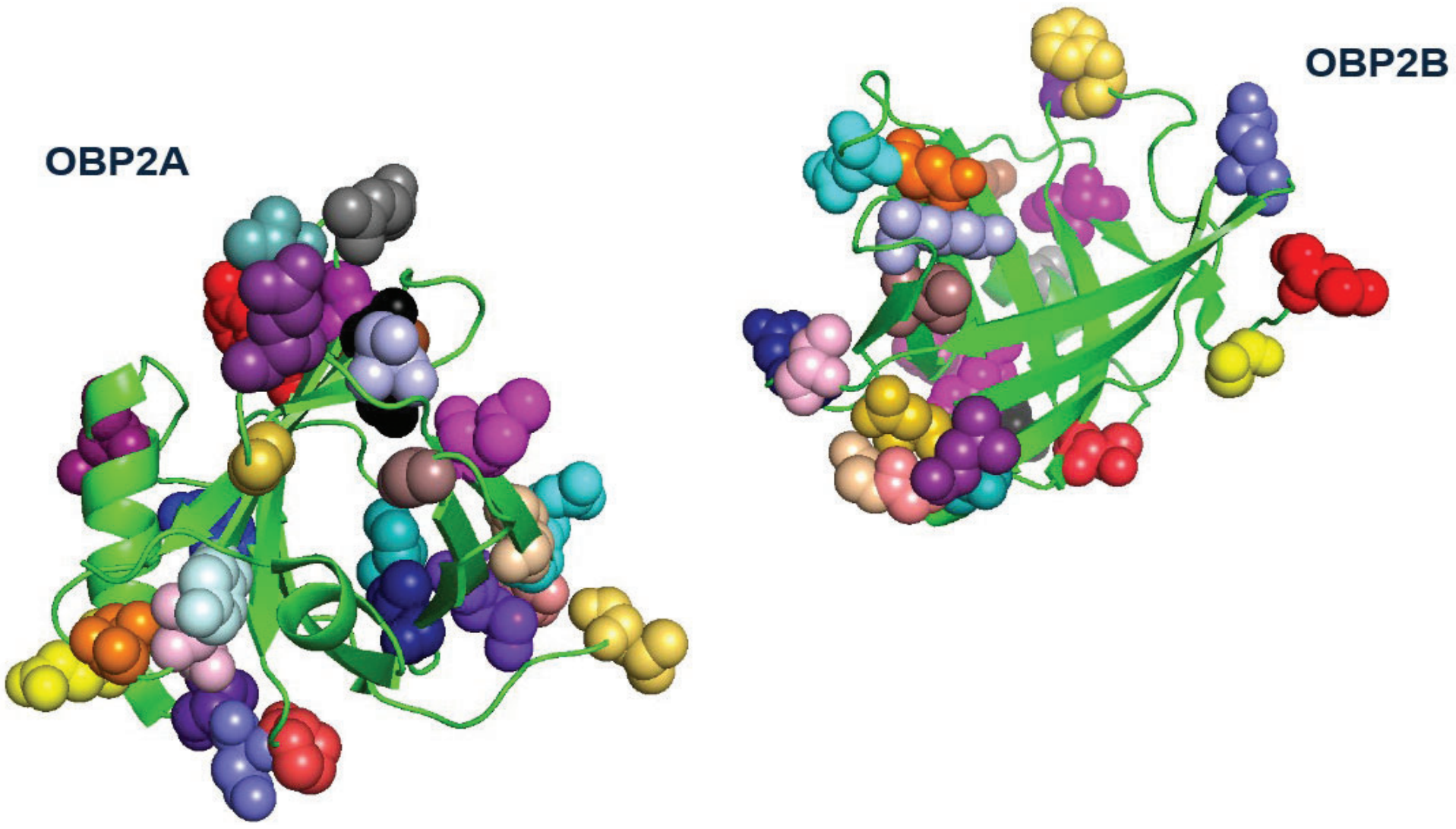

**Figure S12. OBP2A and OBP2B mutations across various cancers.** 44 mutations in OBP2A (left, PDB-ID: 4RUN) and 31 mutations in OBP2B (right, AlphaFold3 predicted structure). It can be seen that the mutations are almost evenly distributed over the two protein structures.

**Table S1 – part 1.** Overview of relevant information on human lipocalins and ORs. The functions are provided as reported in the Human Protein Atlas, GTEx, PaxDB, and PDC.

| Potein | Full name | Size | Gene Position | Links reported to cancer | Reported ligands | Reported functions | Reported pathways (KEGG, GO) | Protein-ID[1] |
|---|---|---|---|---|---|---|---|---|
| OBP2A | Odorant binding protein 2A | ~159 | 9q34.3 | Reported in ovarian cancer, breast cancer | Odorants, small hydrophobic molecules | Odorant transport, potential hormone transport | GO:0005549 (odorant binding) | 4RUN |
| OBP2B | Odorant binding protein 2B | ~164 | 9q34.2 | Emerging links to hormone-related cancers | Odorants, small hydrophobic molecules | Odorant transport, potential link to hormones | GO:0008509 (transporter activity) | - |
| LCN1 | Lipocalin-1 | ~176 | 9q34.3 | Breast cancer | Lipophilic molecules | Pheromone binding, lipid transport | GO:0006869 (lipid transport) | 1XKI; 3EYC |
| LCN2 | Lipocalin 2 | ~198 | 9q34.11 | Reported in renal and breast cancer progression | Iron-loaded siderophores, lipophilic molecules | Iron transport, immune response | KEGG: hsa04974 (protein digestion and absorption), GO:0006879 (cellular iron ion homeostasis) | 1LCP; 4IAW; 5MHH |
| LCN6 | Lipocalin 6 | ~178 | 9q34.3 | No specific links | Unknown | Likely involved in sperm maturation | GO:0007283 (spermatogenesis) | - |
| LCN8 | Lipocalin 8 | ~180 | 9q34.3 | No specific links | Unknown | Potential role in reproduction | GO:0007283 (spermatogenesis) | - |
| LCN9 | Lipocalin 9 | ~180 | 9q34.3 | No specific links | Unknown | Potential role in reproduction | GO:0007283 (spermatogenesis) | - |
| LCN10 | Lipocalin 10 | ~180 | 9q34.3 | Linked to obesity and diabetes | Steroids, fatty acids | Hormonal regulation, metabolism | KEGG: hsa04920 (adipocytokine signaling) | - |
| LCN12 | Lipocalin 12 | ~169 | 9q34.3 | No specific links | Fatty acids, retinoids, and steroids | Lipid metabolism and transport | GO:0008289 (lipid binding) | 2L5P |
| LCN15 | Lipocalin 15 | ~175 | 9q34.3 | Linked to obesity and diabetes | Steroids, fatty acids | Hormonal regulation, metabolism | KEGG: hsa04920 (adipocytokine signaling) | 2XST |
| AMBP | Alpha-1-microglobulin/bikunin precursor | ~160 | 9q32 | Indirectly linked to oral cancer | Salivary proteins | Regulation of saliva viscosity | GO:0050962 (protein binding) | 1BIK; 3QKG; 6EJA; 6EJ9 |
| C8G | Complement component 8, gamma polypeptide | ~200 | 8q11.2 | Linked to obesity and metabolic diseases | Lipids, fatty acids | Lipid metabolism | KEGG: hsa05204 (chemical carcinogenesis) | 2OVA; 1IW2 |
| PAEP | Progestagen-associated endometrial protein | ~125 | 17q24 | Some association with breast cancer | Retinol, vitamin A | Immune response | GO:0045087 (antigen processing) | 4R0B |
| ORM1 | Orosomucoid-1 | ~169 | 9q32 | Linked to various cancers indirectly | Lipids, fatty acids | Inflammatory response | KEGG: hsa04668 (TNF signaling pathway) | 3KQ0 |
| ORM2 | Orosomucoid-2 | ~169 | 9q32 | Linked to obesity and metabolic issues | Lipids, vitamins | Inflammatory response | KEGG: hsa04668 (TNF signaling pathway) | 8QOG; 3APW |
| PTGDS | Prostaglandin D Synthase | ~190 | 9q34.3 | Possible role in prostate cancer | Prostaglandins | Enzyme catalysis, inflammation regulation | KEGG: hsa00590 (arachidonic acid metabolism) | 3O19; 3O22; 8HTA |
| RBP4 | Retinol-binding protein-4, | ~183 | 10q23.3 3 | Diabetes and metabolic disorders (indirect) | Retinol (Vitamin A) | Retinol transport, vision-related | KEGG: hsa04916 (retinol metabolism) | 5NU7; 3FMZ; 6QB; 4O9S; 2WQ9 |
| ApoD | Apolipoprotein D | ~169 | 3q26 | Glioblastoma, breast cancer | Lipids, steroids | Lipid transport, antioxidant activity | GO:0008289 (lipid binding) | 2HZQ; 2HZR |
| ApoM | Apolipoprotein M | ~149 | 6p21 | Possible links to atherosclerosis | Lipids | Lipid transport, immune response | KEGG: hsa04960 (Atherosclerosis) | 2WEX; 2WEW; 2YG2 |
| OR2T5 | Olfactory receptor family 2 subfamily T member 5 | ~315 | 1q44 | - | Odorants | Smell | KEGG: hsa04740 | **Q6IEZ7** |
| OR2T6 | Olfactory receptor family 2 subfamily T member 6 | ~308 | 1q44 | Breast cancer | Galaxolide, odorants | Smell, breast tumorigenesis | KEGG: hsa04740 | **Q8NHC8** |
| OR5B21 | Olfactory receptor family 5 subfamily B member 21 | ~309 | 11q12.1 | Breast cancer | Odor molecules | Smell, breast tumorigenesis | KEGG: hsa04740 | **A6NL26** |
| OR51E1 | Olfactory receptor family 51 subfamily E member 1 | ~318 | 11p15.4 | Prostate cancer, Colorectal cancer | Propionate, Butyrate,Isovaleric acid, Nonanoic acid, Decanoic acid, β-Ionone, Hexanoic acid, Octanoic acid | Odorant receptor, chemosensory receptor for metabolites | KEGG: hsa04740 | **Q8TCB6** |
| OR51E2 | Olfactory receptor family 51 subfamily E member 2 | ~188 -237 | 11p15.4 | Prostate cancer, Breast cancer | β-Ionone, Acetate, Propionate, Nonanoic acid, Butyrate, Octanoic acid, Decanoic acid, Lactate, Methyl decanoate, 4-Ethylphenol | Metabolite sensing, Neuroendocrine transition | KEGG: hsa04740 | **Q9H255** |

[1]When no protein crystal structure is available in the PDB, we provide uniprot-IDs (highlighted in bold).

**Table S1 – part 2.** Overview of relevant information on human lipocalins and ORs. The functions are provided as reported in the Human Protein Atlas, GTEx, PaxDB, and PDC.

| Protein | RNA - Normal (HPA) | RNA - normal (GTEx/HPA) | RNA - cancer (HPA) | Protein- normal (PaxDb) | Protein- cancer (PDC) |
|---|---|---|---|---|---|
| OBP2A | Reproductive organs, nose, skin | **YES** | YES | YES | ovarian |
| OBP2B | Reproductive organs, nose, skin | **YES** | YES | YES | breast |
| LCN1 | Reproductive organs, Brain, Endocrine tissues, Respiratory system, Proximal digestive tract, Pancreas, Muscle tissues, Bone marrow & Lymphoid tissues | **YES** | YES | YES | ovarian, castric |
| LCN2 | Kidney, liver, neutrophils | **YES** | YES | **YES** | brain, ovarian, breast, colon, lung, uterine, pancreatic |
| LCN6 | Testis-specific | **YES** | YES | YES | - |
| LCN8 | Testis-specific | **YES** | YES | YES | - |
| LCN9 | Testis-specific | YES | NO | **YES** | - |
| LCN10 | Various tissues, particularly in liver | **YES** | YES | **YES** | - |
| LCN12 | Skeletal muscle and heart muscle | **YES** | YES | YES | - |
| LCN15 | Adipose tissue, liver | **YES** | YES | **YES** | colon, uterine |
| AMBP | Salivary glands | **YES** | YES | **YES** | prostate, ovarian, uterine, brain, breast, colon |
| C8G | Liver, adipose tissue | **-** | - | - | - |
| PAEP | Placenta | **-** | - | - | - |
| ORM1 | Liver | **YES** | YES | **YES** | brain, colon, ovarian, breast |
| ORM2 | Liver | **YES** | YES | YES | brain, breast, colon, gastric |
| PTGDS | Brain, male genital tract | - | YES | YES | brain, breast, colon |
| RBP4 | Liver, adipose tissue | - | - | - | - |
| ApoD | Brain, breast tissue | - | - | - | - |
| ApoM | Various tissues, including liver | - | - | - | - |
| OR2T5 | Salivary gland | YES | NO | YES | - |
| OR2T6 | Eye | YES | NO | YES | - |
| OR5B21 | Testis | YES | YES | YES | - |
| OR51E1 | Endocrine tissues, Proximal digestive tract, Gastrointestinal tract, Liver & Gallbladder, Pancreas, Kidney & Urinary bladder, Male tissues, Female tissues, Muscle tissues, Connective & Soft tissue, Skin, Bone marrow & Lymphoid tissues | **YES** | YES | **YES** | - |
| OR51E2 | Gastrointestinal tract, Male tissues | **YES** | YES | YES | - |

**Table S2.** List of the different residues between OBP2A/2B as obtained through MSA (with a total 17 residues that differ between OBP2A and OBP2B). The residue numbers correspond to the crystal structure of OBP2A (PDB-ID: 4RUN). The four residues in blue bold change their properties from hydrophobic to hydrophilic, while the orange one changes from hydrophilic to hydrophobic (see also visualization in Figure S8). These changes are mostly likely to impact the binding specificity between the two proteins.

| Residue numbers | OBP2A | OBP2B |
|---|---|---|
| 46 | N - Hydrophilic | K - Hydrophilic |
| 75 | F - Hydrophobic | Y - Hydrophobic |
| 84 | I - Hydrophobic | M - Hydrophobic |
| 91 | G - Hydrophilic | R - Hydrophilic |
| 92 | T - Hydrophilic | R - Hydrophilic |
| 94 | D - Hydrophilic | H - Hydrophilic |
| 96 | V - Hydrophobic | I - Hydrophobic |
| 103 | R - Hydrophilic | H - Hydrophilic |
| 104 | R - Hydrophilic | H - Hydrophilic |
| **108** | **R - Hydrophilic** | **L - Hydrophobic** |
| **109** | **Y - Hydrophobic** | **H - Hydrophilic** |
| **118** | **P - Hydrophobic** | **H - Hydrophilic** |
| 119 | N - Hydrophilic | D - Hydrophilic |
| **122** | **L - Hydrophobic** | **R - Hydrophilic** |
| 134 | H - Hydrophilic | R - Hydrophilic |
| **144** | **M - Hydrophobic** | **T - Hydrophilic** |
| 153 | L - Hydrophobic | P - Hydrophobic |

**Table S3. Known biomarkers for different cancer types.** List of known protein markers used in differential gene expression analysis for different cancer types. These biomarkers served as reference points to assess the relationship between OBP overexpression and significant genes in each cancer within the volcano plots generated in this review.

| *Cancer type* | *Known protein markers* |
|---|---|
| Ovarian cancer | HLA-DRA, CLDN4, CLDN3 |
| Breast cancer | ERBB2, KRT19, KRT18 |
| Lung cancer | SPAG5, KIF23, RAD54L |
| Endometrial cancer | ESPL1, PRAME, PTTG1 |
| Colorectal cancer | KLK6, SFTA2, LEMD1 |
| Prostate cancer | EPHA10, HPN, SLC4A2 |

**Table S4. Filter Parameters for Samples from GTEx and TCGA Database.** This table presents the filter parameters used for selecting healthy (GTEx) and tumor (TCGA) samples for each cancer type. Six cancer types were investigated for differential gene expression analysis. For GTEx, filter parameters included primary site and primary tissue, while for TCGA, the primary site was used, encompassing all histological types.

| *Cancer type* | *GTEx filter parameters* | | *TCGA filter parameters* | *Number of Samples* | |
|---|---|---|---|---|---|
| | *paraPrimarySiteGTEx* | *paraPrimaryTissueGTEx* | *paraPrimarySiteTCGA* | *GTEx* | *TCGA* |
| *Ovarian cancer* | Ovary | Ovary | Ovary | 87 | 418 |
| *Breast cancer* | Breast | Breast - Mammary Tissue | Breast | 177 | 1092 |
| *Uterine cancer* | Uterus | Uterus | Uterus | 77 | 57 |
| *Lung cancer* | Lung | Lung | Lung | 286 | 1011 |
| *Colorectal cancer* | Colon | Colon | Colon | 302 | 287 |
| *Prostate cancer* | Prostate | Prostate | Prostate | 99 | 494 |

**Table S5: Background Correction for Mutation Cases in Human Lipocalins, OBP2A, and OBP2B in Melanoma and Endometrial Cancer.** This table presents the expected number of mutation samples due to random chance for all human lipocalin genes, OBP2A, and OBP2B in melanoma and endometrial cancer. TCGA data for each cancer type were used, and mutation samples were counted based on missense mutations. The average mutation counts per sample were calculated.

| Gene | Cancer type | Data | # Mutation Samples | *Average of mutation counts* | Human genome nucleotides |
|---|---|---|---|---|---|
| All human lipocalins | Melanoma | Skin Cutaneous Melanoma (TCGA, PanCancer Atlas) | 440 | 740.5 | 3200000000 |
| OBP2A | Endometrial Cancer | Uterine Corpus Endometrial Carcinoma (TCGA, PanCancer Atlas) | 515 | 1046 | 3200000000 |
| OBP2B | Melanoma | Skin Cutaneous Melanoma (TCGA, PanCancer Atlas) | 440 | 740.5 | 3200000000 |

| Gene | Lipocalin exon nucleotides | Chance hitting any nt in the genome | Chance hitting lipocalin exons | #Samples in the TCGA | Correction |
|---|---|---|---|---|---|
| All human lipocalins | 17075 | 0.00000023140625 | 0.003951261719 | 444 | 1.754360203 |
| OBP2A | 702 | 0.000000326875 | 0.00022946625 | 586 | 0.1344672225 |
| OBP2B | 962 | 0.00000023140625 | 0.0002226128125 | 444 | 0.09884008875 |